\documentclass[journal]{IEEEtran}                                                        
\usepackage{graphicx}
\usepackage{amsmath}
\usepackage{textcomp}
\usepackage{float}
\usepackage{amsmath,amssymb}
\usepackage{wrapfig}
\usepackage{bm}
\usepackage{color}
\usepackage{breqn}
\usepackage{scalerel,stackengine}

\usepackage{balance}

\title{\Huge
{Nonlinear Characterization of Tissue Viscoelasticity with Acoustoelastic Attenuation of Shear-Waves}
}

\author{Bhaskara R. Chintada, Richard Rau, Orcun Goksel

\thanks{B.R. Chintada, R. Rau, and O. Goksel are with the Computer-assisted Applications in Medicine (CAiM) group, Swiss Federal Institute of Technology (ETH), Zurich, Switzerland. (email: ogoksel@ethz.ch)

}}
\begin{document}

\maketitle
\thispagestyle{empty}

%%%%%%%%%%%%%%%%%%%%%%%%%%%%%%%%%%%%%%%%%%%%%%%%%%%%%%%%%%%%%%%%%%%%%%%%%%%%%%%%
\begin{abstract}
Shear-wave elastography (SWE) measures shear-wave speed (SWS), which is related to the underlying shear modulus of soft tissue. 
SWE methods generally assume that soft tissue viscoelasticity is independent of mechanical loading, however, soft tissues are known to have viscoelasticity changing nonlinearly with pre-compression. 
Hence, characterization by SWS alone is insufficient, where nonlinear properties can be seen as confounders but may also be utilized as additional bio-markers.
The viscoelastic nature of a medium is fully characterized by its storage and loss moduli, which are related to SWS and shear-wave attenuation (SWA).
%Additionally, the nonlinear viscoelastic characteristics can be quantified and parameterized based on SWS and SWA values with respect to applied strains based on .
In this work, we study SWA characteristics as a function of applied strain to measure nonlinear viscoelastic parameters in soft tissues. 
For this purpose, we apply incremental quasi-static compression on the samples while measuring SWS and SWA, from which we derive storage and loss moduli to estimate nonlinear viscoelastic parameters as a function of applied strain using acoustoelasticity (AE) theory.  
Results from tissue-mimicking phantoms with varying oil percentages and \textit{\textit{ex-vivo}} porcine liver experiments demonstrate the feasibility of the proposed approach. 
In both these experiments, SWA was observed to decrease with applied strain.
For 10\% compression in \textit{ex-vivo} livers, shear-wave attenuation decreased on average by 28\% (93\,Np/m), while SWS increased on average by 20\% (0.26\,m/s).


\end{abstract}
%%%%%%%%%%%%%%%%%%%%%%%%%%%%%%%%%%%%%%%%%%%%%%%%%%%%%%%%%%%%%%%%%%%%%%%%%%%%%%%%
\section{INTRODUCTION}
Shear-wave elastography (SWE) is a noninvasive imaging technique that maps shear-wave speed (SWS) in a tissue. 
Shear-waves are induced by acoustic radiation force and their propagation is observed using ultrasound imaging to capture lateral shear-wave travel speed~\cite{sarvazyan1998shear}, which is related to the underlying shear modulus of tissue. 
SWS measurements have been used in many clinical applications including diagnosis of diseases in liver, breast, and kidney~\cite{sarvazyan2011elasticity,deffieux2015investigating,tanter2008quantitative}. 
SWS however has shown poor specificity in breast cancer diagnosis~\cite{kim2015false} and in the early-stage diagnosis of nonalcoholic fatty liver disease (NAFLD).  
For the latter, a major reason is the fat accumulation with steatosis, which may cause a reduction in SWS, acting as confounder for the expected SWS increase with simultaneously advancing fibrosis~\cite{poynard2013liver,chintada2019acoustoelasticity}.

For the further understanding and diagnosis of pathological changes in soft tissues, additional biomechanical markers would be greatly beneficial. 
Although SWE assumes that soft tissues are linear and elastic, they are inherently nonlinear and viscoelastic~\cite{fung2013biomechanics}. 
Several bio-markers have been proposed to characterize tissue nonlinearity and viscoelasticity to complement SWS measurements. 
For instance, in~\cite{barry2012shear} dispersion of SWS (i.e., variation with frequency) was shown to correlate with the degree of viscosity induced by steatosis. 
In general, storage $G'$ and loss $G''$ moduli can together fully characterize viscoelastic behaviour of a medium. 
These moduli can be derived from SWS and shear-wave attenuation (SWA) (i.e., decay of shear-wave amplitude with distance)~\cite{catheline2004measurement}. In~\cite{garteiser2012mr}, it was found for magnetic resonance elastography (MRE) of human liver tumors that the loss modulus provides better differential diagnosis between benign and malignant tumors than the storage modulus. 
With this motivation, several studies have since been conducted to estimate SWA accurately~\cite{nenadic2016attenuation,budelli2017diffraction,bernard2016frequency,kijanka2019two}. 
In~\cite{nenadic2016attenuation}, SWA was measured at a particular frequency from its full width at half-maximum (FWHM) in k-space from the 2D-FFT of shear-wave propagation vs time profiles. 
Dependency of FWHM on length of spatial observation window (i.e.\ the segment length of the propagation-distance) used to compute 2D-FFT was investigated and accounted for with a model in~\cite{rouze2017accounting}. 
SWA as the rate of shear-wave amplitude decay based on a cylindrical wavefront assumption was studied in~\cite{budelli2017diffraction,kazemirad2016ultrasound}. In~\cite{bernard2016frequency}, a frequency-shift (FS) method was introduced to measure SWA by fitting a model to the amplitude spectrum of shear-wave displacements at different propagation distances.
Similarly, a two-point frequency-shift (2p-FS) method was developed in~\cite{kijanka2019two} by fitting a model to the amplitude spectrum of shear-wave displacements at any two propagation distances. 

Nonlinear characteristics of soft tissues using SWE have been studied by measuring the change in SWS values at various incremental stresses using the theory of acoustoelasticity (AE) in quasi-incompressible media~\cite{latorre2012quantitative,otesteanu2017quantification,bernal2016vivo,bayat2019acoustoelasticity,aristizabal2018application}. In~\cite{latorre2012quantitative,otesteanu2017quantification} nonlinear properties of liver samples were investigated. 
In~\cite{bernal2016vivo}, it was shown that benign and malignant lesions in the breast exhibited different nonlinear shear modulus.
In~\cite{bayat2019acoustoelasticity}, changes in the nonlinear shear modulus were studied in bladder samples before and after treatment with formalin.
In~\cite{aristizabal2018application}, the effect of compression area and nonlinear characteristics of kidney samples was investigated for progressive and regressive compression. 
In~\cite{jiang2015characterization} and ~\cite{jiang2015measuring}, the Demiray-Fung hyperelastic model was employed to quantify nonlinear properties of \textit{ex-vivo} porcine liver tissue and \textit{ex-vivo} brain tissue, respectively.
Frequency variations of the nonlinear shear modulus were studied recently in~\cite{otesteanu2019spectral} using SWS dispersion characteristics in \textit{ex-vivo} porcine liver samples.
Note that in all of the above mentioned nonlinear characterization studies, the shear storage modulus is approximated as the shear modulus by neglecting the attenuation term, under the assumption that soft tissues are elastic.
%Accordingly, SWS $v$ is related to shear modulus $G$ using the empirical relationship \mbox{$G=\rho {v}^2$}, where $\rho$ is the tissue density. 
The storage modulus with the attenuation term, the loss modulus and SWA characteristics as a function of applied stress/strain have so far not been investigated, which is essential in order to fully quantify the nonlinear viscoelastic characteristics of soft-tissues under the influence of stress (compression). 
This is highly relevant given that the loss modulus is potentially superior to the storage modulus in differential diagnosis of benign and malignant tumors in the human liver~\cite{garteiser2012mr}. 
Furthermore, SWA is has shown to be effective in segregation of transplanted livers with acute rejection vs.\ no rejection~\cite{nenadic2016attenuation}. 
Such a characterization would not only yield additional bio-markers for diagnosis, but would also help in taking into account any potential confounding effects that may be introduced, e.g.\ by compression, during the measurement of SWS and SWA. 

In this paper, based on AE theory and using acoustic radiation force (ARF) induced shear-waves, we study SWA changes with applied strain levels for characterizing tissue nonlinear viscoelastic properties, in particular storage and loss moduli. 
%The study is carried out on tissue mimicking oil-gelatin phantoms to mimic statosis progression as well as \textit{ex-vivo} porcine liver samples.
%To the best of our knowledge, this is the first work that studies SWA characteristics as a function of applied strain and quantifies nonlinear viscoelastic mechanical parameters based on AE theory using acoustic radiation force induced shear-waves.
%%%%%%%%%%%%%%%%%%%%%%%%%%%%%%%%%%%%%%%%%%%%%%%%%%%%%%%%%%%%%%%%%%%%%%%%%%%%%%%%%%%%%%%%%%%
\section{Material and Methods}
% Note: Data processing steps are summarized in 'Data_processing_summary.png', I request you to take help of it while commenting in Material and methods section. 
\subsection{Data acquisition}
For this study, we employ shear-waves induced via ARF using the supersonic shear-wave imaging technique~\cite{bercoff2004supersonic}, which generates a cylindrical shear-wave front.
We use three subsequent high-intensity ARF pushes of 200\,$\mu$s duration at three axially separated foci with a separation of 5\,mm in depth. 
Shear-wave propagation is then tracked utilizing ultrafast ultrasound imaging at 10\,k frames-per-second and coherent compounding of three angled plane waves at (-8\textdegree{},0\textdegree{},8\textdegree{}), using a moving average filter of last three frames.
This acquisition sequence was programmed in a research ultrasound machine (Verasonics, Seattle, WA, USA) with a 128-element linear-array transducer (Philips, ATL L7-4) operated at 5\,MHz center frequency. 

\begin{figure} 
\includegraphics[width=\columnwidth]{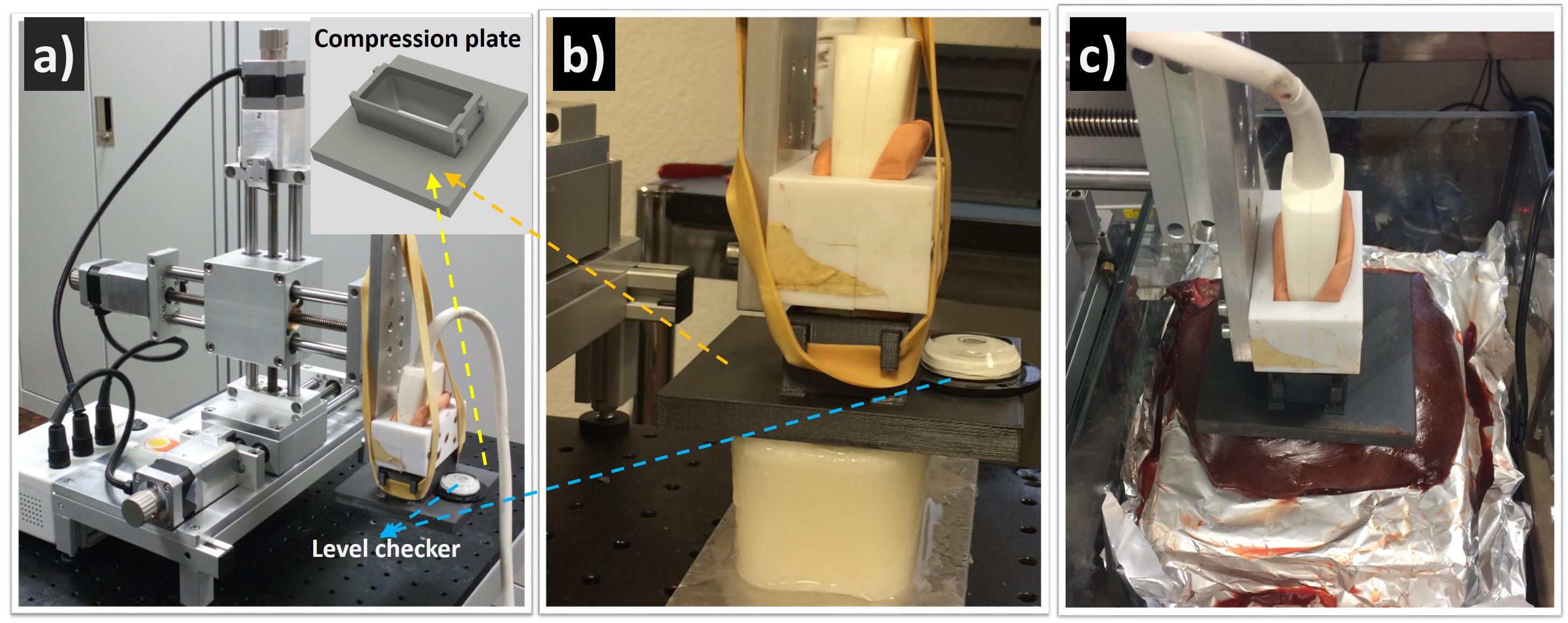}
\caption{Overview of the experimental setup with the automated motorized linear-stage (a) characterizing a tissue mimicking oil gelatin phantom (b) and an \textit{ex vivo} porcine liver sample (c). The phantoms and liver samples are placed on a rigid parallel plate and incremental quasi-static stress was applied with a compression plate.}
\label{Exp_setup}
\end{figure}

Shear-wave acoustoelasticity experiments were conducted by applying compression using a motorized three-axis linear stage in 1\% increments up to 10\% strain.
The ultrasound probe, attached to a positioning stage, was used for inducing shear-waves via ARF and imaging the resulting shear-waves. 
To ensure uni-axial compression on phantoms, a custom designed 3D printed compression plate of dimensions 100\,mm $\times$ 100\,mm (with a small window for imaging through) was attached on the front of the ultrasound probe, as shown in Fig.\,\ref{Exp_setup}a.
From the shear-wave in-phase and quadrature (IQ) signal acquired at each compression step, the shear-wave propagation is tracked using a 2D Loupas autocorrelation  method~\cite{loupas1995axial} after directionally filtering the left and right propagating waves~\cite{manduca2003spatio} along the depth axis $z$, leading to displacement profiles $u(r,t)$, where $r$ is the lateral distance to the ARF source, and $t$ the time. 

\subsection{Shear-wave propagation in viscoelastic medium}
\label{SWProp}
ARF-induced shear-wave in a viscoelastic medium is commonly modeled as a cylindrical wave originating from an infinite source line~\cite{deffieux2008shear}, defined in the frequency domain as:
\begin{subequations}
    \begin{align}
        \label{eq:1a}
        \hat u(r,\omega) & \approx \frac{i}{4} \sqrt{\frac{2}{\pi \hat{k}(\omega)r}}e^{(-i\hat{k}(\omega)r +\frac{\pi}{4})}\\
        \label{eq:1b}
        \mathrm{with\quad} \hat{k}(\omega) &= \frac{\omega}{c_p(\omega)}-i\alpha(\omega) 
    \end{align}
\end{subequations}
where $\hat u(r,\omega)$ is the Fourier transform of the shear wave displacement field $u(r,t)$ polarized along depth (i.e. $z$-axis), $\omega$ is the angular frequency, $\hat{k}(\omega)$ is the complex wave number, $c_p(\omega)$ is the phase velocity, and $\alpha(\omega)$ is the frequency dependent attenuation.

As seen in Eq.\,\eqref{eq:1a}, the shear-wave amplitude decreases as a function of the propagation distance $r$ due to medium viscous absorption and diffraction. 
Indeed, the shear-wave phase velocity and attenuation can be derived based on the Fourier domain spectrum in Eq.\,\eqref{eq:1a}.
Taking the natural logarithm of the amplitude $|{\hat u(r,\omega)}|$ and separating the phase terms to equate to phase angle $\angle\hat u(r,\omega)$ yield, respectively: 
\begin{subequations}
    \label{eq:SWProp}
    \begin{align}
    \label{eq:SWPropa}
    \ln (|{\hat u(r,\omega)}|) &= {-\alpha(\omega) r} - \frac{1}{2}\ln(r)- \ln(A(\omega)),\\
    \label{eq:SWPropb}
    \angle\hat u(r,\omega) &= \Big[{\frac{\omega}{c_p(\omega)}\Big]r-\theta _0}\,.
    \end{align}
\end{subequations} 
Where $A(\omega)$ is a constant dependent on the frequency $\omega$, and $\theta _0$ is the phase angle at the wave source (ARF push point).
Eq.\,\eqref{eq:SWPropa} allows to estimate frequency-dependent attenuation $\alpha(\omega)$ also taking into account of the diffraction term $\frac{1}{2}\ln(r)$, while 
Eq.\,\eqref{eq:SWPropb} helps to estimate phase velocity $c_p(\omega)$. % A schematic describing the estimation of  $c_p(\omega)$ and SWA $\alpha(\omega)$ can be seen in Fig.\,. \ref{SWS_SWA_estimation}.

For a linear and isotropic viscoelastic medium, the complex shear modulus $\hat{G}(\omega)$ is defined by the storage $G'(\omega)$ and loss $G''(\omega)$ moduli and is related to the complex shear-wave number $\hat{k}$ through
\begin{equation}  \label{eq:complex_shear_wave_number}
    G'(\omega)+ iG''(\omega) = \hat{G}(\omega) = \frac{\rho\omega^2}{\hat{k}^2(\omega)},
\end{equation}
where $\rho$ represents the tissue density. 
Substituting Eq.\,\eqref{eq:complex_shear_wave_number} in Eq.\,\eqref{eq:1b} allows to derive the shear storage and loss moduli as
\begin{subequations}
    \label{eq:complex_mod}
    \begin{align}
        \label{eq:storage_mod}
        G'(\omega) &= \rho \omega^2 \frac{\big(\frac{\omega}{c_p(\omega)}\big)^2-\alpha(\omega)^2}{\Big(\big(\frac{\omega}{c_p(\omega)}\big)^2+\alpha(\omega)^2\Big)^2}, \\
        \label{eq:loss_mod}
        G''(\omega) &= 2\rho \omega^2 \frac{\big(\frac{\omega}{c_p(\omega)}\big)\alpha(\omega)}{\Big(\big(\frac{\omega}{c_p(\omega)}\big)^2+\alpha(\omega)^2\Big)^2}. 
    \end{align}
\end{subequations}

In conventional SWE, shear storage modulus is often referred simply as shear modulus and it is commonly estimated from Eq.\,\eqref{eq:storage_mod} by neglecting shear-wave attenuation $\alpha(\omega)$ as
\begin{equation}  
    \label{eq:G}
    G(\omega) = \rho c^2_{p}(\omega).
\end{equation}

%Theoretical storage $G'(\omega)$ and loss $G''(\omega)$ moduli values are shown in Fig. \ref{G_1_G_11_theory}
%
%\begin{figure}
%\includegraphics[width=0.5\textwidth]{G_1_G_11_theory.png}
%\caption{Storage $G'$ and loss $G''$ moduli computed over a range of $c_p(\omega)$ values varying from 1-3 m/s and $\alpha(\omega)$ values varying from 10-500 Np/m at frequency 300Hz using the empirical relationship mentioned in Eqs.\, \ref{eq:storage_mod} \& \ref{eq:loss_mod} respectively}. 
%\label{G_1_G_11_theory}
%\end{figure}
%
%over a typical range of $c_p(\omega)$ and $\alpha(\omega)$ values in tissue mimicking phantoms and soft tissues.

%from 1 to 3 m/s , and range of $\alpha(\omega)$ values from 10 to 500 Np/m at 300Hz.

\subsection{Acoustoelasticity in viscoelastic medium}
\label{AE_theory}
%Nonlinear coefficients for an incompressible isotropic medium were obtained from the expansion of strain energy density ($W$) up to fourth-order \cite{hamilton2004separation} as 
%\begin{equation}
%      W &= \mu I_2 + \frac{A}{3}I_3 + DI_2^2,
%      \label{eq:strain_density}
%\end{equation}
%where $I_2$ and $I_3$ are second and third Lagrangian strain invariants, $\mu$ is a Lam\'e parameter, $A$ is the third order elastic coefficient as defined by Landau-Lifschitz, and $D$ is the fourth order elastic coefficient as defined by Hamilton, in this study we neglect fourth order terms since the shear wave amplitudes are very small similar to the studies \cite{gennisson2007acoustoelasticity,}.

The influence of uniaxial compression on the speed of acoustic waves in elastic, isotropic, homogenous and nearly compressible media are studied using the theory of AE.
For a shear-wave polarized along the axis of deformation, Gennisson \emph{et\,al.}~\cite{gennisson2007acoustoelasticity} approximated a linear dependence of shear modulus on uniaxial strain $\epsilon$ as follows:
%by considering the first two terms of strain energy expansion up to fourth order (i.e. eq. \eqref{eq:strain_density}) 

\begin{equation}\label{eq:AE_elastic}
     \rho v_{\mathrm{g}}^2 = \mu_0+ \Big(\frac{A}{4}\Big)\epsilon,
\end{equation}
where $\rho$ is the tissue density, $v_g$ the shear-wave group speed, $\mu_0$ the shear modulus at stress-free (resting) condition (i.e., $\epsilon = 0$), and $A$ the third-order elastic constant.

%For larger deformation of tissues, Destrade et al. \cite{destrade2010third} derived the quadratic dependence of shear modulus on uniaxial strain considering all the terms in eq. \eqref{eq:strain_density} as
%\begin{equation}
%     \rho v_{\mathrm{s}}^2 &= \mu_0+ (\frac{A}{4})\epsilon + (2 \mu_0+ A + 3D) \epsilon^2.
%        \label{eq:AE_elastic_b},
%\end{equation}
% where $D$ is the 4th order elastic constant. 

However, in a viscoelastic medium, the shear-wave velocity becomes complex and frequency-dependent, and hence the shear modulus also becomes complex and frequency-dependent. 
Substituting Eq.\,\eqref{eq:complex_shear_wave_number} in Eq.\,\eqref{eq:AE_elastic} results in
%
%\begin{subequations}
%    \begin{align}
%        \hat{G}(\omega) &= \hat{\mu_0(\omega)} + (\frac{\hat A(\omega)}{4})\epsilon, \label{eq:AE_viscoelastic}\\
%        \hat{G(\omega)} &= \hat{\mu_0(\omega)} + (\frac{\hat A(\omega)}{4})\epsilon+(2\hat\mu_0(\omega)+\hat A(\omega)+3\hat D(\omega))\epsilon^2.
%        \label{eq:AE_viscoelastic_b}
%    \end{align}
% \end{subequations}
%
\begin{equation}  
       \label{eq:AE_viscoelastic}
       \hat{G}(\omega) = \hat\mu_0(\omega) + \Big(\frac{\hat A(\omega)}{4}\Big)\epsilon,
\end{equation}
where $\hat\mu_0(\omega)=\mu_0'(\omega)+i\mu_0''(\omega)$ is the complex shear modulus at stress-free condition, and $\hat{A}(\omega)=A'(\omega)+i A''(\omega)$ is the 3rd-order complex nonlinearity coefficient, composed of an elastic $A'(\omega)$ and viscous $A''(\omega)$ nonlinearity components.

%and $\hat{D}(\omega)= D'(\omega)+i D''(\omega)$ is the fourth order nonlinear complex coefficient.
%In this study, we will use eq. \eqref{eq:AE_viscoelastic} \& \eqref{eq:AE_viscoelastic_b} for investigating the nonlinear viscoelastic coefficients $\hat{A}(\omega)=A'(\omega)+i A''$ and $\hat{D}(\omega)= D'(\omega)+i D''(\omega)$. 3rd order nonlinear complex coefficient $\hat{A}(\omega)=A'(\omega)+i A''$ can be estimated using either of the equations, while eq. \eqref{eq:AE_viscoelastic_b} is used for estimating $\hat{D}(\omega)= D'(\omega)+i D''(\omega)$. A schematic describing the process can be seen in Fig. \ref{AE_procedure}.

\begin{figure*} 
\includegraphics[width=1\textwidth]{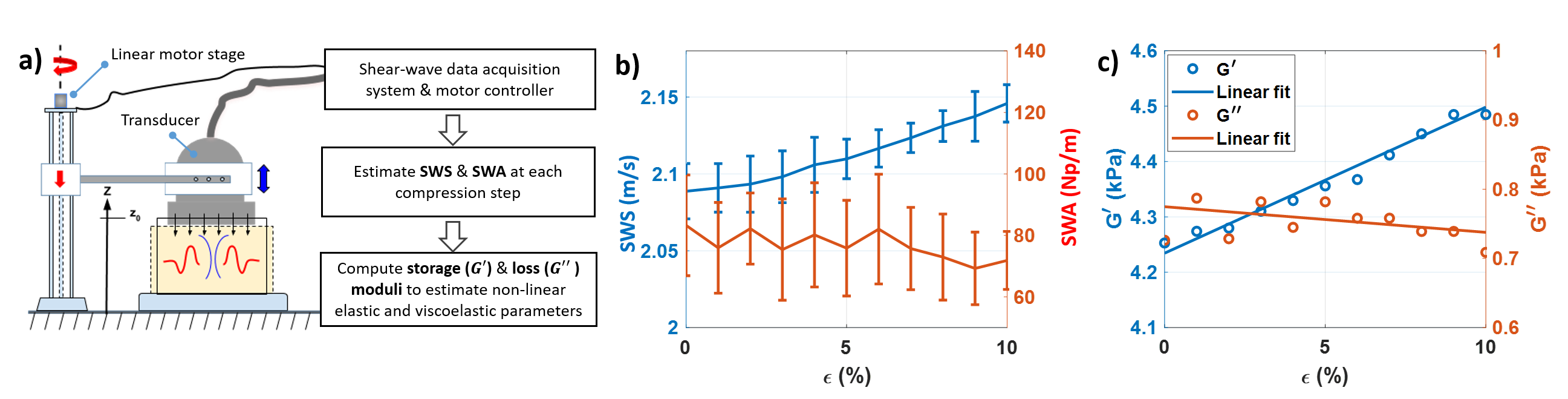}
\caption{Schematic illustrating the overall procedure of determining the nonlinear viscoelastic parameters. Shear-wave data is acquired at each compression step by applying stress on the targeted medium placed on a rigid surface using an ultrasound transducer connected to a linear motor stage setup (a).
Shear-wave speed and attenuation is computed over a range of frequencies at each compression step to first estimate storage ($G'$) and loss ($G''$) moduli with Eq.\,\eqref{eq:complex_mod} and then extract nonlinear parameters using Eq.\,\eqref{eq:AE_viscoelastic}.
For e.g., shear wave speed and attenuation at frequency 300\,Hz with applied strain are plotted in (b) and the derived $G'$ and $G''$ are shown in (c) along with its linear-fit to estimate in nonlinear viscoelastic parameters $A'$ and $A''$, respectively.
%At every compression step shear-wave phase velocity $c_{p}(\omega)$ and attenuation $\alpha(\omega)$ is computed to derive storage and loss modulus ($G' (\omega)$, & $G''(\omega)$). Nonlinear viscoelastic parameters ($A'$ and $A''$) at frequency 300\,Hz are extracted by fitting Eq.\,\eqref{eq:AE_viscoelastic}.
} 
\label{AE_procedure} 
\end{figure*}

In this study, we use Eq.\,\eqref{eq:AE_viscoelastic} to estimate a complex nonlinearity coefficient from multiple complex shear moduli measurements at different strains. 
A schematic of our experimental process can be seen in Fig.\,\ref{AE_procedure}. 
%
%In order to compare nonlinear viscoelastic parameters with nonlinear elastic parameters, we estimate $A$ and $D$ using eqs. \eqref{eq:AE_elastic} \& \eqref{eq:AE_viscoelastic_b}.
%
In practice, at each sample compression step, the phase velocity $c_{p}(\omega)$ and attenuation $\alpha(\omega)$ are computed based on Eq.\,\eqref{eq:SWProp} using the Fourier transform of tracked shear-waves (the detailed procedure will be explained later in the section.\,\ref{implementation}).
From $c_{p}(\omega)$ and $\alpha(\omega)$, the complex shear modulus $\hat{G}(\omega)$ is derived using Eq.\,\eqref{eq:complex_mod}.
Finally, the linear model in Eq.\,\eqref{eq:AE_viscoelastic} is fitted separately to the real and imaginary moduli to estimate the nonlinear elastic $A'(\omega)$ and viscous $A''(\omega)$ constants (see Fig.\,\ref{AE_procedure}c). 
Note that even the estimation of the \emph{elastic} component of the 3rd-order nonlinearity constant would be affected whether or not shear-wave attenuation is taken into account.
In~\cite{otesteanu2019spectral}, shear-wave attenuation was neglected by estimating an elastic constant $A(\omega)$ from $G(\omega)$.
This is equivalent to substituting shear-wave phase velocity $c_{p}(\omega)$ instead of shear wave group speed $v_g$ in Eq.\,\eqref{eq:AE_elastic}.
We compare this approach with $A'(\omega)$ computed using actual storage modulus $G'(\omega)$ based on our derivation herein by taking attenuation into account.
%In order to quantify the difference between computing 3rd order elastic constant from $G$ by neglecting the shear-wave attenuation (i.e. $A(\omega)$) and from {$G\prime$} by considering the shear-wave attenuation (i.e. $A'(\omega)$). We computed $A(\omega)$ by substituting shear-wave phase velocity $c_{p}(\omega)$ instead of shear wave group speed $v_g$ in eq. \eqref{eq:AE_elastic}, similar to the study in \cite{otesteanu2019spectral}. 
We also computed magnitude of complex 3rd order viscoelastic constant (i.e. $|\hat{A}(\omega)|$) by substituting the magnitude of $\hat G(\omega)$ (i.e. $|\hat G(\omega)|=\sqrt{G'(\omega)^2+ G''(\omega)^2}$) in Eq.\,\eqref{eq:AE_viscoelastic} to quantify total nonlinearity of the targeted medium.
%EQ
In summary, we compute $A$, $A'$, $A''$, and $|\hat{A}(\omega)|$ using following equations, respectively: 
%\begin{eqnarray}
\begin{subequations}
\label{eq:all_nl_params}
\begin{align}
    A(\omega)\, \epsilon &= 4\left( G(\omega) - \mu_0(\omega)\right) \label{eq:A_estimate}\,,\\
    A'(\omega)\, \epsilon &= 4\left( G'(\omega) - \mu_0'(\omega)\right) \label{eq:A_prime_estimate}\,,\\
    A''(\omega)\, \epsilon &= 4\left( G''(\omega) - \mu_0''(\omega)\right) \label{eq:A_double_prime_estimate}\,,\\
    |\hat A(\omega)|\, \epsilon &= 4\left( |\hat{G}(\omega)| - |\hat{\mu_0}(\omega)| \right) \label{eq:A_magnitude_estimate}\,.
\end{align}
\end{subequations}
    
%\end{eqnarray}
At any given frequency $\omega$, from multiple readings of $\epsilon$ under different compressions, right-hand-side parameters are first found as described earlier.
We then solve for any left-hand-side nonlinearity parameter in least-squares sense, i.e.\ a line-fitting problem. 

\subsection{Tissue mimicking phantom experiments}
Four tissue mimicking gelatin phantoms were prepared with increasing castor oil content of \{0\%, 5\%, 10\%, 15\%\} to model increasing viscosity similarly to the studies in~\cite{barry2012shear,kazemirad2016ultrasound}, herein we call these phantoms \{\#1..\#4\}.
%steatosis progression, while keeping the gelatin content fixed at 8\%.
%These phantoms, called herein \{\#1..\#4\}, then have increasing viscosity similarly to the studies in~\cite{barry2012shear,kazemirad2016ultrasound}. 
For the phantom preparation, we follow a procedure similar to~\cite{madsen2006anthropomorphic}: We pre-heated distilled water to $\approx$70\textdegree{}C before mixing in predetermined percentages of gelatin (G2500, Sigma-Aldrich chemical, USA) and castor oil (259853, Sigma-Aldrich chemical, USA). 
Anionic surfactant (Pril Original, Germany) of concentration 4\,cc/L was added to make oil droplets sufficiently small for castor oil to mix homogeneously with gelatin solution. 
After the mixture is cooled down to $\approx$50\textdegree{}C, 1\% of cellulose (S5504, Sigma-Aldrich chemical, USA) was added to provide scattering, essential for shear-wave tracking. 
Phantom compositions are tabulated in Tab.~\ref{Ph_ingredients_Tab1}.
\begin{table}
  \caption{Compositions of the four tissue mimicking phantoms, in percentages to total mixture by weight.}
  \label{Ph_ingredients_Tab1} 
  \begin{center}
    \begin{tabular}{l||c|c|c}
      \textbf{Phantoms} & \textbf{Gelatin [\%] } & \textbf{Castor Oil [\%]} & \textbf{Surfactant [\%]}\\
      \hline
      \#1 & \phantom{0}8& \phantom{0}0 & 0\phantom{.2}  \\
      \#2 & \phantom{0}8& \phantom{0}5 & 0.2 \\
      \#3 & \phantom{0}8& 10 & 0.2 \\
      \#4 & \phantom{0}8& 15 & 0.2 \\
    \end{tabular}
  \end{center}
\end{table}
The mixture was then let to cool down further and, while still in a semi-liquid state, was poured in a 100$\times$40$\times$50\,mm$^3$ mould. 
Until pouring, the mixture was stirred at a low speed throughout the preparation process, in order to avoid oil separation from gelatin and to ensure a homogeneous scatterer distribution. 
The phantom mixtures were then refrigerated at 4\textdegree{}C for 18\,h. 
Prior to data acquisition, the phantoms were removed from their moulds and kept at room temperature for 6\,h for a uniform temperature distribution. 
Shear-wave data was acquired by placing the phantoms on a hard flat surface, while applying compression with the compression plate parallel to this surface, as shown in Fig.\,\ref{Exp_setup}b.

To improve SNR, shear-wave displacement fields over 11 acquisitions at each compression step were averaged prior to further processing steps.
Compression with 0.5\,mm steps up to 5\,mm were applied, corresponding to 1\% increments with the given phantom thickness.

To take care of the bias, that may arise due to changes in experimental conditions, three cycles of compression were performed on each phantom with 15\,mins waiting in-between the measurements, such that the phantoms could recover from applied deformation during previous data acquisition.
During this waiting period, the phantom were kept in a plastic storage box to prevent drying.

%At least two data acquisitions are used in computing the nonlinear parameters. 

\subsection{\textit{Ex-vivo} liver experiments}
To test the feasibility of the proposed measurements on actual tissue and to study the mechanical characteristics of liver tissue, an \textit{ex-vivo} study was performed on four porcine livers, fresh on the same day they were brought from the local slaughterhouse. 
The livers were placed in a flat container as shown in Fig.\,\ref{Exp_setup}c, and shear-wave data were collected for compression steps of 0.3\,mm up to 3\,mm for sections of livers that are roughly 30\,mm in thickness. 
Similarly to phantom experiments, 3 sets of measurements were collected for each liver with 20\,mins waiting in-between measurements for the liver parenchyma to recover. 
The liver samples were kept in 0.9\% saline water between the experiments to avoid dehydration (also to prevent the intrinsic water squeezing out of the porous liver parenchyma under its own weight). 
Similarly to~\cite{otesteanu2019spectral}, to improve SNR, displacements fields over 30 acquisitions were averaged prior to further processing steps.

\subsection{Implementation} \label{implementation}
\begin{figure*}  
    \centering
    \includegraphics[width=\textwidth]{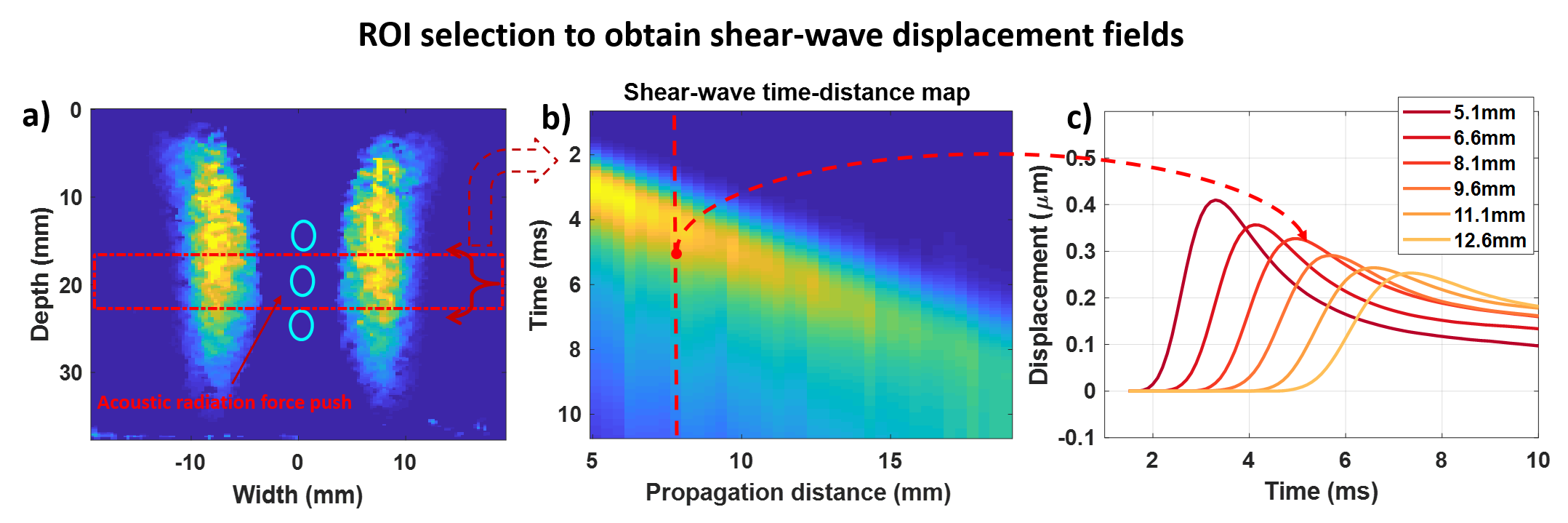}
    \caption{(a) Shear-wave displacement field at 1 ms after the generation of radiation force pushes. (b) Shear-wave propagation distance-time maps are obtained by averaging  over a window of 5\,mm around the central push. (c) Shear-wave displacement profiles obtained at different propagation distances are plotted.}
    \label{ROI_selection}
\end{figure*}

Using supersonic approach, we induced shear-waves displacement fields $u(z,r,t)$  by three consecutive ARF pushes with 5\,mm separation, as shown in Fig.\,\ref{ROI_selection}a. 
To increase SNR, we averaged axial displacements from an axial region-of-interest (black box in Fig.\,\ref{ROI_selection}a).
To ensure a planar wavefront, we used a 5\,mm axial window around the center push.
This then reduces the displacement field $u(z,r,t)$ in the ROI to $u(r,t)=\sum_{\mathrm{ROI}}u(z,r,t)$.

%Averaging over ROI with window height between 2\,mm and 10\,mm was suggested in~\cite{brum2014vivo} and~\cite{deffieux2009shear} to improve the signal-to-noise ratio (SNR) of shear-wave displacement fields, in this work, ROI with height of 5\,mm was chosen around the center push for both tissue mimicking phantoms and ex-vivo liver experiments as illustrated in Fig.\,\ref{SWS_SWA_estimation}.

\begin{figure*} 
\includegraphics[width=\textwidth]{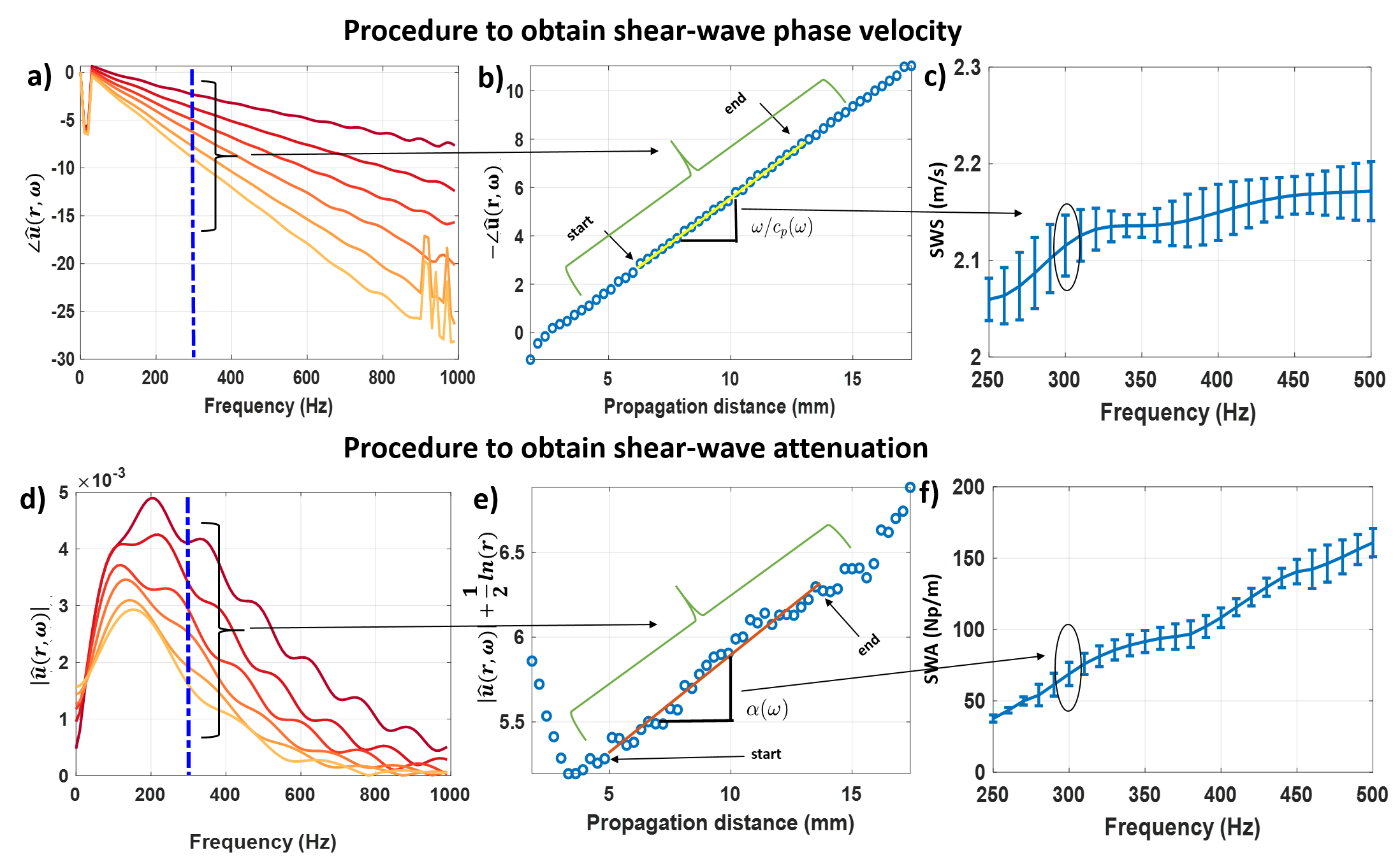} 
\caption{Schematic describing the procedure to estimate the shear-wave phase velocity and attenuation dispersion profiles. 
1D temporal Fourier transform is performed at different propagation distances to obtain phase profiles $\angle\hat u(r,\omega)$ (a) and amplitude spectrum profiles $|\hat u_{z}(r,\omega)|$ (d).
Shear-wave phase velocity $c_{p}(\omega)$ is obtained from reciprocal of the slope of varying $-\angle\hat u(r,\omega)$ profiles (b) while shear-wave attenuation $\alpha(\omega)$ is obtained from the slope of $-|\hat u(r,\omega)|+ \frac{1}{2}\ln(r)$ profiles (e) at different lateral distances ${r_{i=1 .  .n}}$.
The resulted SWS and SWA values at the 300\,Hz frequency marked in dot-dashed line in (a) and (d) represent the datapoints in the ellipses in (c) and (f), respectively.
The same procedure is repeated for a range of frequencies and $c_{p}(\omega)$ is plotted in (c) while $\alpha(\omega)$ is plotted in (f).
These computation are performed on the shear-wave data-set of Phantom \#3.}
\label{SWS_SWA_estimation}  
\end{figure*}

The 1D Fourier transform $\hat{u}(r,\omega)$ of $u(r,t)$ at different propagation distances ${r_{i=1 .  .n}}$ are computed to estimate shear-wave phase velocity $c_p(\omega)$ and attenuation $\alpha(\omega)$.
Typical shear-wave phase angles $\angle(\hat u(r,\omega)$ and amplitude spectrum $|\hat u(r,\omega)|$ at different propagation distances are shown in Fig.\,\ref{SWS_SWA_estimation}a/d, respectively.
In the following step, $c_p(\omega)$ and $\alpha(\omega)$ were obtained using the phase and amplitude profiles of $\hat{u}(r,\omega)$ according to Eq.\,\eqref{eq:SWProp} using 'polyfit' function in \mbox{MATLAB}.
The shear-wave phase velocities $c_p(\omega)$ were estimated from the slope of varying $\angle(\hat u(r,\omega)$ over a selected window of propagation distances as illustrated in Fig.\,\ref{SWS_SWA_estimation}b. 
The frequency-dependent shear-wave attenuation $\alpha(\omega)$ is estimated as negative slope of varying shear-wave amplitudes with diffraction correction (i.e. $\|\hat u(r,\omega)\|+ \frac{1}{2}\ln(r)$) as illustrated in Fig.\,\ref{SWS_SWA_estimation}e.

\begin{figure} 
\centering
\includegraphics[width=.90\columnwidth]{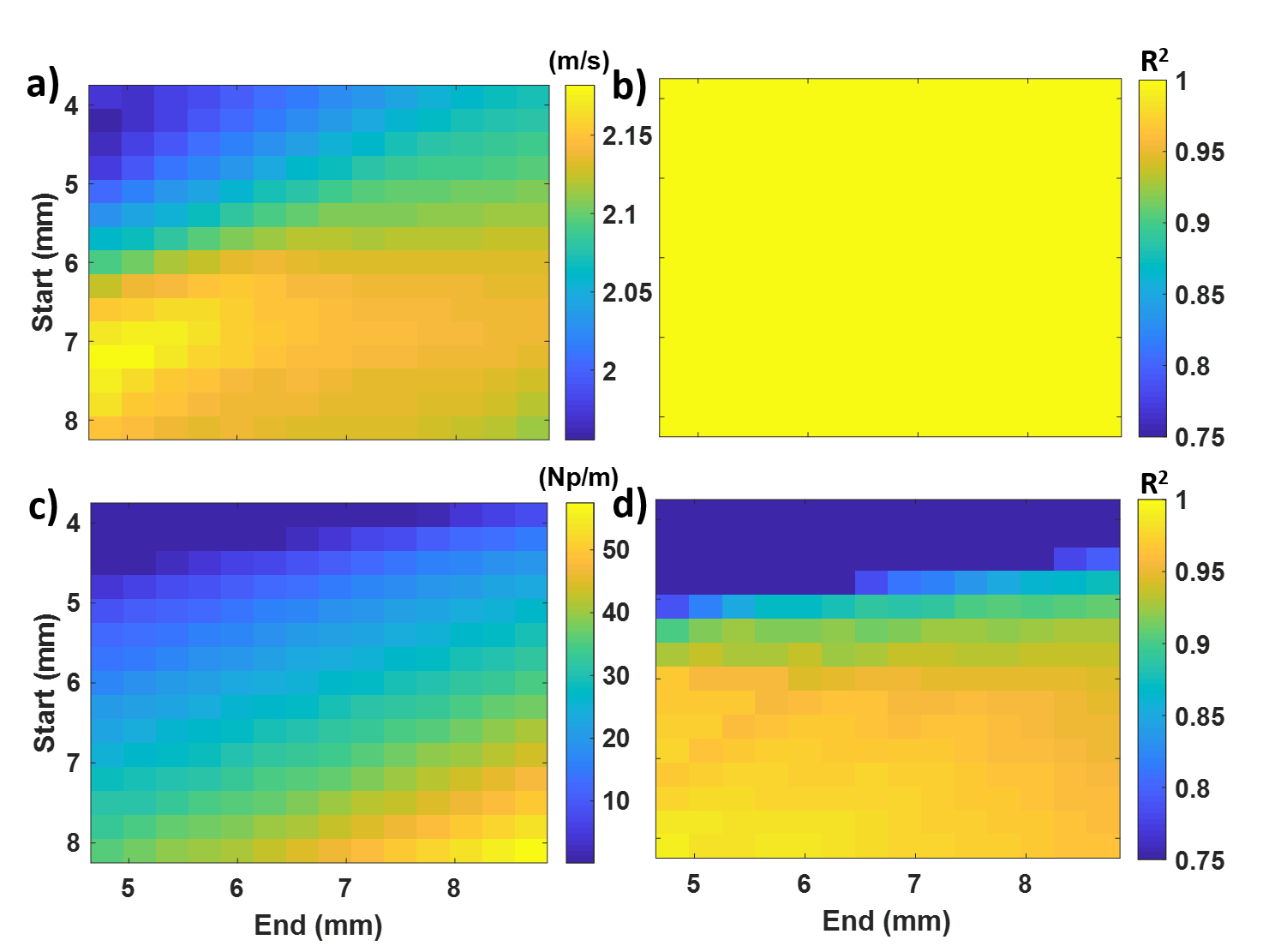}
\caption{Shear-wave phase velocity (a) and its R$^2$ values of fit (b), shear-wave attenuation measurements (c) and its R$^2$ values of fit (d) for various combination of start and end propagation distance positions.}
\label{SWS_SWA_computation_method}
\end{figure}

Calculations were performed for various combinations of start and end propagation distances to derive the best estimate of $c_p(\omega)$ and $\alpha(\omega)$ (see Fig.\,\ref{SWS_SWA_computation_method}a,c).
For this purpose, while computing $c_{p}(\omega)$ and $\alpha(\omega)$, corresponding coefficient of determination (R-square) of fit was computed to evaluate goodness of the fit and plotted in Fig.\,\ref{SWS_SWA_computation_method}b,d corresponding to estimates in Fig.\,\ref{SWS_SWA_computation_method}a,c respectively.
Only the estimates with R-square greater than 0.95 were considered for phantoms and greater than 0.90 for \textit{ex-vivo} porcine livers for further processing.
Note that this computational procedure is different from~\cite{budelli2017diffraction} and~\cite{lipman2016feasibility}.
In~\cite{lipman2016feasibility} shear-wave attenuation values were computed by fixing first propagation distance and searching for the optimal end propagation distance with minimum standard error of fit, while in~\cite{budelli2017diffraction} full transducer aperture size [0-15\,mm] was considered.
The selection for start and endpoint is crucial, because near field shear-wave variations and faster decaying higher shear-wave frequency components create aberrations in the near field and in the far shear-wave propagation distances respectively (see Fig.\,\ref{SWS_SWA_estimation}e).
At the in-between propagation distances, the shear-wave amplitudes are uniform due to formation of full shear-wave displacement profiles.
Considering these limitations on amplitude profiles, it is difficult to accurately estimate shear-wave attenuation and depending on the chosen start and end point of propagation distances, estimated attenuation values are either over or underestimated. 
Hence, computing $c_p(\omega)$ and $\alpha(\omega)$ values over varying start and end lateral distances and taking the average of estimates having \mbox{goodness-of-fit} greater than specific threshold are expected to yield more accurate $\alpha(\omega)$.
This approach is similar to the  FS and 2P-FS methods in~\cite{bernard2016frequency} and~\cite{kijanka2019two}, respectively.
%Similar characteristics of shear-wave amplitude profiles also discussed in \cite{budelli2017diffraction}, FS \cite{bernard2016frequency} and 2P-FS \cite{kijanka2019two} methods. For the same reason, FS and 2P-FS method computations were performed over various first signal positions and later segment lengths.
These limitations are also valid in case of the phase profiles, i.e. the $c_p(\omega)$ derivation, even though less prominent compared to amplitude profiles, as can be seen in Fig.\,\ref{SWS_SWA_estimation}e.
Hence, the estimation of $c_p(\omega)$ is based on the same procedure.

The resulted values are marked in Fig.\,\ref{SWS_SWA_estimation}c for $c_p(\omega)$ and in Fig.\,\ref{SWS_SWA_estimation}f for $\alpha(\omega)$, the same process was repeated for a range of frequencies, for the SWE data acquired at different strain levels, and for different compression cycles.
In the following step, storage $G'(\omega)$ and loss $G''(\omega)$ moduli were derived from averaged $c_p(\omega)$ and $\alpha(\omega)$ over different compression cycles using the relationship mentioned in Eq.\,\eqref{eq:AE_viscoelastic}.
To fit the functions in Eq.\,\eqref{eq:all_nl_params} and derive the nonlinear viscoelastic parameters from $G'(\omega)$ and $G''(\omega)$, we used the MATLAB least-squares fitting function \textit{fit}.
In order to quantify the goodness of the fit, we have used the metric root-mean-square error (RMSE) similar to~\cite{aristizabal2018application} and considered only the outcomes having \mbox{RMSE$<$0.25\,kPa}. 
%With the AE experiments, $c_p(\omega)$ and $\alpha(\omega)$ obtained at multiple compression levels can be used to extract the nonlinear parameters of viscoelastic medium (i.e. $ A', A'', D'$ and $D''$). These parameters are computed by fitting the corresponding models, eq. \eqref{eq:AE_viscoelastic} to estimate $ A'$ and $A''$ and eq. \eqref{eq:AE_viscoelastic_b} to estimate $D'$ and $D''$ were used. The nonlinear parameters obtained from the assumption that medium is pure elastic from eqs. \eqref{eq:AE_elastic} \& \eqref{eq:AE_elastic_b} are denoted by A and D.
%%%%%%%%%%%%%%%%%%%%%%%%%%%%%%%%%%%%%%%%%%%%%%%%%%%%%%%%%%%%%%%%%%%%%%%%%%%%%%%%%%%%%%%%%%%%%%%%%%%%
\section{Results}
\subsection{Phantom results}
\label{Ph_results}
\begin{figure*}  
    \centering
    \includegraphics[width=\textwidth]{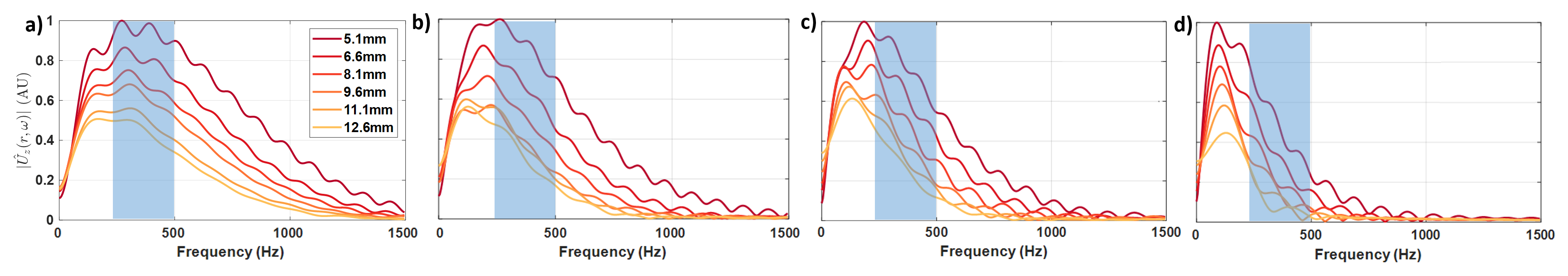}
    \caption{1D temporal Fourier transform of shear-wave displacement signals at different propagation distances at stress-free condition, i.e.\ $\epsilon\approx 0$ for phantoms \{\#1-\#4\} mentioned in Tab.~\ref{Ph_ingredients_Tab1}.}
    \label{Ph_FFT_profiles}
\end{figure*}

Normalized amplitude spectra calculated at propagation distances $r=\{5.1, 6.6, 8.1, 9.6, 11.1, 12.6\}$\,mm at stress-free condition are shown for each phantom in Fig.\,\ref{Ph_FFT_profiles}.
Since the viscosity increases with increasing oil percentage, the amplitude spectrum gets narrower from phantoms \#1 to \#4, with a meaningful SNR not left above 500\,Hz in \#4, whereas the peak amplitude not being reached until $\approx$250\,Hz in \#1.
This common frequency range [250, 500]\,Hz is thus chosen to comparatively analyze all phantoms, following the similar convention in~\cite{parker2018analysis}.

\begin{figure} 
\includegraphics[width=\columnwidth]{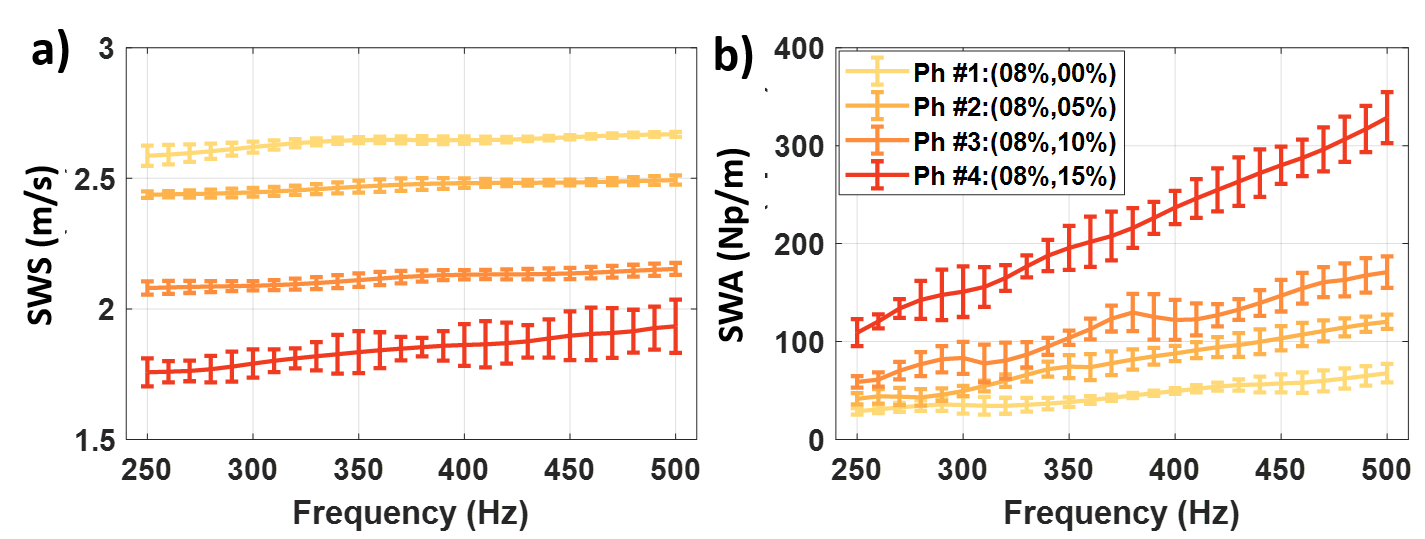}
\caption{Shear-wave phase velocity $c_p(\omega)$ (a), attenuation $\alpha(\omega)$ (b) as a function of frequency, measured for the phantoms {\#1-\#4} at stress-free condition, i.e.\ $\epsilon\approx 0$.
The error bars correspond to the standard deviation of the respective estimates having R-square value greater than 0.95.}
 \label{Ph_SWS_SWA_no_stress}
\end{figure}

\begin{figure}  
\includegraphics[width=\columnwidth]{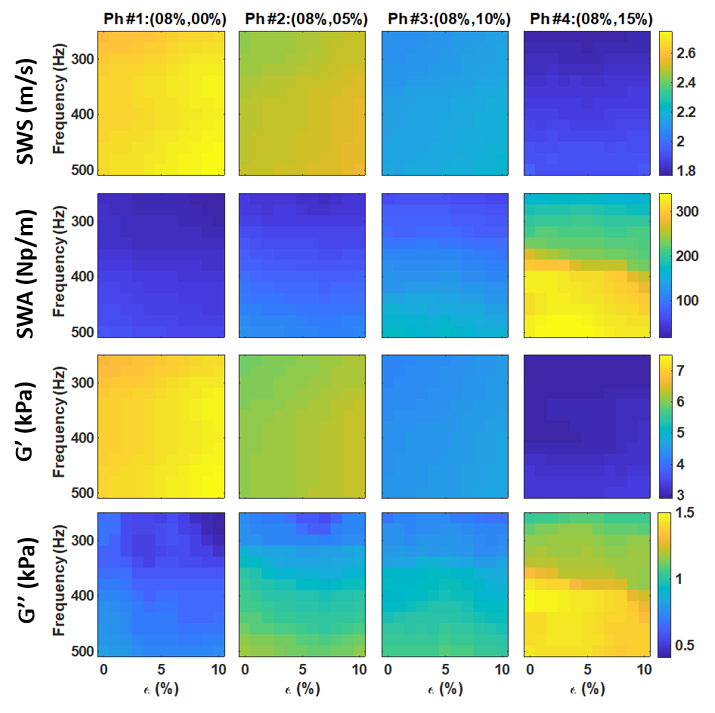}
\caption{Shear-wave phase velocity (1st row), attenuation (2nd row), storage $(G')$ (3rd row) and loss $(G'')$ moduli (bottom row) as a function of frequencies and applied strain levels for phantoms \#1-\#4.}
\label{Ph_SWS_SWA_G_1_G_11}
\end{figure}

Fig.\,\ref{Ph_SWS_SWA_no_stress} shows our estimated frequency-dependent shear-wave phase velocity and attenuation values at stress-free condition, i.e.\ $\epsilon\approx 0$.
The error bars correspond to the standard deviation of the respective estimates having R-square value greater than 0.95.
At any given frequency it is observed that, with increased oil percentage (i.e.\ from phantom \#1 to \#4), the shear-wave phase velocity decreases considerably, whereas the attenuation increases.
Indeed, phantoms {\#1-\#4} respectively exhibit a shear-wave velocity dispersion of \{0.01, 0.02, 0.03, 0.08\}\,[m/s]/100\,Hz, and a shear-wave attenuation dispersion of \{20.5, 24.5, 43.0, 84.5\}\,[Np/m]/100\,Hz, computed from the profiles in Fig.\,\ref{Ph_SWS_SWA_no_stress} using a linear model. 
%
%Storage and loss moduli are increasing with oil percentage and showed dispersion of 84.3, 98.6, 99.2 and 173.1 [Pa]\100\,Hz in storage modulus and dispersion of 69.0, 110.2, 130.8 and 260 [pa]\100\Hz in loss modulus. It can be observed that dispersion in loss modulus is more prominent against the dispersion in storage modulus.  
%It can be observed that $G'(\omega)$ to $G''(\omega)$ are decreasing with increasing oil percentages. For e.g. at frequency 300\,Hz $G'$ to $G''$ ratio is 10.8, 7.4, 5.6 and 3.03 for phantoms \{#1..#4\} respectively 
%
For all the phantoms, the estimated shear-wave phase velocity and attenuation, as well as the storage and loss moduli derived from the former two, are shown in Fig.\,\ref{Ph_SWS_SWA_G_1_G_11} as a function of frequencies and applied strain.
The maps look homogeneous due to the large color-scale ranging across the phantoms as well as the relatively lower nonlinearity of gelatin phantoms (compared to liver tissue presented later).

\begin{figure*}
\includegraphics[width=\textwidth]{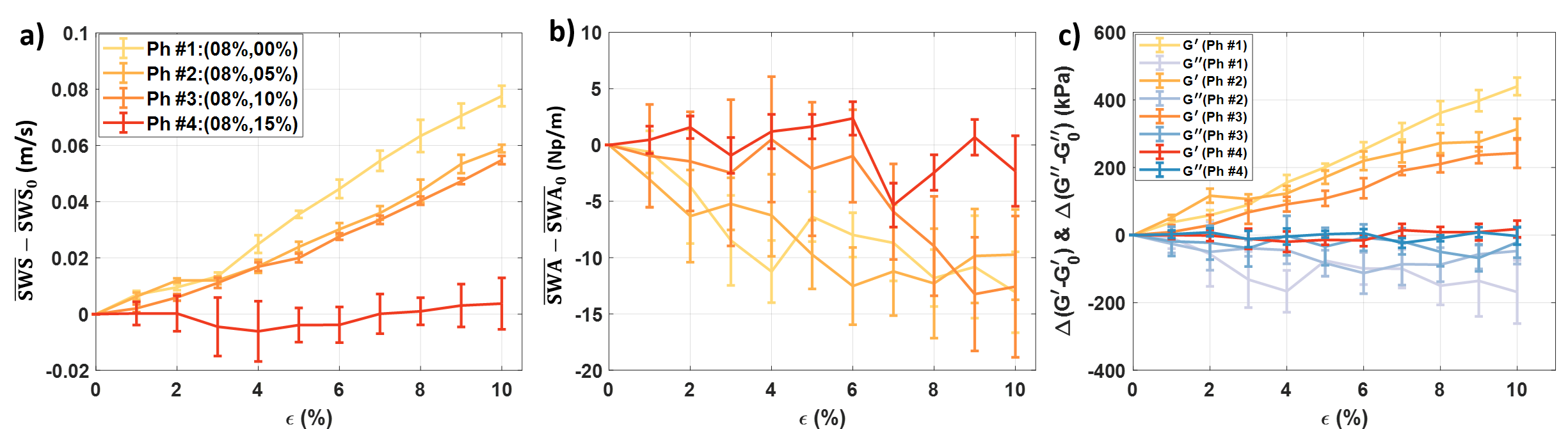}
\caption{Frequency-Averaged shear-wave phase velocity (a), attenuation (b), storage and loss moduli (c) with respect to frequencies on the corresponding maps in Fig.\,\ref{Ph_SWS_SWA_G_1_G_11} and offset by their nominal values, respectively, SWS\textsubscript{0}, SWA\textsubscript{0}, $G'$\textsubscript{0}, and $G''$\textsubscript{0} at stress-free state are plotted as function of applied strain ($\epsilon$), the error bars represent the standard deviation of these values.}
 \label{Ph_SWS_SWA_wrt_compression}
\end{figure*}

In order to visualize absolute changes of corresponding parameters as a function of applied strain, in Fig.\,\ref{Ph_SWS_SWA_wrt_compression} we present the shear-wave phase velocity, attenuation, storage and loss moduli as averaged with respect to frequency axis and offset by their nominal values, respectively, SWS\textsubscript{0}, SWA\textsubscript{0}, $G'$\textsubscript{0}, and $G''$\textsubscript{0} at stress-free state.
It is observed that, having 10\% strain applied in phantoms \#1-\#4, respectively, shear-wave phase velocity changes by +2.9\%, +2.5\%, +2.4\%, and +0.06\%, while attenuation changes by -27.7\%, -16.1\%, -10.0\%, and -0.8\%, indicating an increase in phase velocity whereas a decrease in attenuation with applied strain.
As expected from the relatively small nonlinearity of these phantoms, shear storage $G'$ and loss $G''$ moduli as functions of applied strain resemble the trends of shear-wave velocity and attenuation in Fig.\,\ref{Ph_SWS_SWA_wrt_compression}a,b respectively.

\begin{figure} 
\includegraphics[width=\columnwidth]{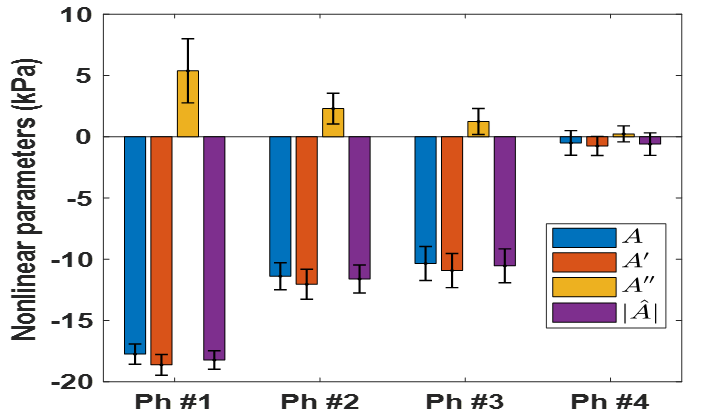}
\caption{Frequency-averaged elastic $A$ and viscoelastic nonlinear parameters $A'$, $A''$, and its magnitude $|\hat{A}|$ are extracted from phantoms \#1-\#4 that were subjected to compression of 0.5\,mm steps for a total of 5\,mm ($\approx$ 10\%). The errorbars indicate the standard deviation across frequencies.}
\label{Ph_NL_params}
\end{figure}

As a biomechanical marker, the nonlinearity parameter at a selected frequency (range) can be used. Alternatively, we report herein frequency-averaged values of these parameter estimates, for further analysis and interpretation. 
Frequency-averaged values together with their standard deviations are presented visually in Fig.\,\ref{Ph_NL_params} and also tabulated in Tab.~\ref{Ph_results_Tab}.
It can be seen that the nonlinear elastic parameters $A$ and $A'$ and nonlinear viscous parameter $A''$ are decreasing exponentially with increasing castor oil percentage.
%suggests the possibility of using these parameters as new bio-marker to diagnose steatosis stages in NAFLD.

The absolute value of the nonlinear elastic parameter $A$ without considering the attenuation term is 5\%, 6\%, 6\% and 13\% lower than true nonlinear elastic parameter $A'$ measured considering the attenuation term for phantoms \mbox{\#1-\#4}, suggesting the importance of attenuation term while computing the shear storage modulus.
The standard deviation of $A$ and its percentages with respect to its mean values are $\pm$5.3\%, $\pm$9.5\%, $\pm$11.5\% and $\pm$164.4\% 
%0.96 (05.3\%), 1.06 (09.5\%), 1.17 (11.5\%), and 0.97 (164.4\%)
for phantoms \#1-\#4 respectively, suggests an increasing variation of $A$ with respect to frequency as oil percentage increases.
A similar trend can also be observed in nonlinear parameters $A'$ and $A''$.

%All frequency averaged mechanical parameters for all the phantoms investigated are presented in Tab.~\ref{Ph_results_Tab}, together with their standard deviations. 
%The ratio of $\mu_'$ to $\mu_0''$ is 9.7, 6.4, 5.4 and 2.9 for phantoms \{#1..#4\} respectively.
%
\begin{table*}
  \caption{Measured nonlinear elastic and viscoelastic parameters ($\pm$ standard deviations) for the phantoms used in this study}
  \label{Ph_results_Tab} 
  \begin{center}
    \begin{tabular}{l||c|c||c|c||c}
      \textbf{Phantom} & \textbf{$\mu_0$} & \textbf{$A$} & \textbf{$\hat{\mu}=\mu_0'+i\mu_0''$} & \textbf{$\hat{A}=A'+iA''$} & \textbf{$|\hat{A}|$}\\ 
     &\textbf{{[kPa]}} &\textbf{{[kPa]}} &\textbf{{[kPa]}} &\textbf{{[kPa]}} & \textbf{{[kPa]}}\\ 
      \hline
      \#1 &  6.94 $\pm$ 0.12 & -17.63 $\pm$ 0.96 & (6.88 $\pm$ 0.12) + i (0.71 $\pm$ 0.06)  & (-18.48 $\pm$ 1.05) + i (5.38 $\pm$ 2.62) & -17.92 $\pm$ 0.98\\
      \#2 &  5.81 $\pm$ 0.14 & -11.13 $\pm$ 1.06 & (5.71 $\pm$ 0.11) + i (0.89 $\pm$ 0.15)  & (-11.99 $\pm$ 1.21) + i (2.29 $\pm$ 1.25) & -11.52 $\pm$ 1.10 \\
      \#3 &  4.79 $\pm$ 0.15 & -10.18 $\pm$ 1.17 & (4.67 $\pm$ 0.11) + i (0.87 $\pm$ 0.15)  & (-10.76 $\pm$ 1.34) + i (1.24 $\pm$ 1.06) & -10.36 $\pm$ 1.22 \\
      \#4 &  3.37 $\pm$ 0.22 & -00.59 $\pm$ 0.97 & (3.10 $\pm$ 0.14) + i (1.07 $\pm$ 0.20)  & (-00.79 $\pm$ 0.77) + i (0.23 $\pm$ 0.65) & -00.66 $\pm$ 0.90 \\
    \end{tabular}
  \end{center}
\end{table*}

\subsection{\textit{Ex-vivo} liver experiments}
\begin{figure*} 
    \centering
    \includegraphics[width=\textwidth]{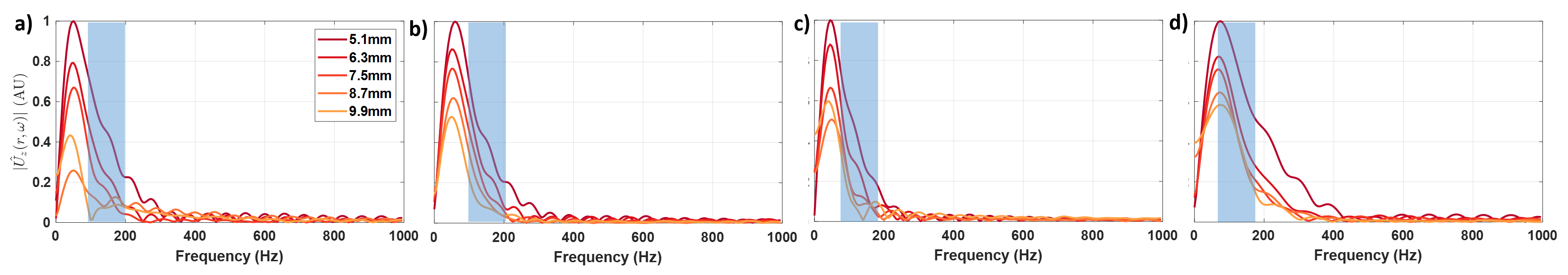}
    \caption{1D temporal Fourier transform of shear-wave of displacement fields at different propagation distances at stress-free condition, i.e.\ $\epsilon\approx 0$ for \textit{ex-vivo} porcine livers.}
    \label{Livers_FT_profiles}
\end{figure*}

The feasibility of our method was next tested on four \textit{ex-vivo} porcine livers subjected to 10\% maximum strain.
The normalized amplitude spectra calculated at propagation distances \mbox{$r=\{5.1, 6.3, 7.5, 8.7, 9.9\}$}\,mm at stress-free condition are shown for each liver in Fig.\,\ref{Livers_FT_profiles}.
Porcine liver shear-wave spectra is seen to be much narrower compared to that of phantoms, due to the higher intrinsic shear-wave attenuation. 
For a comparative analysis between the livers, a common frequency range [100, 200]\,Hz was chosen, given the maximum peak and maximum sufficient signal ranges, as described earlier for the phantoms.

\begin{figure} 
\includegraphics[width=\columnwidth]{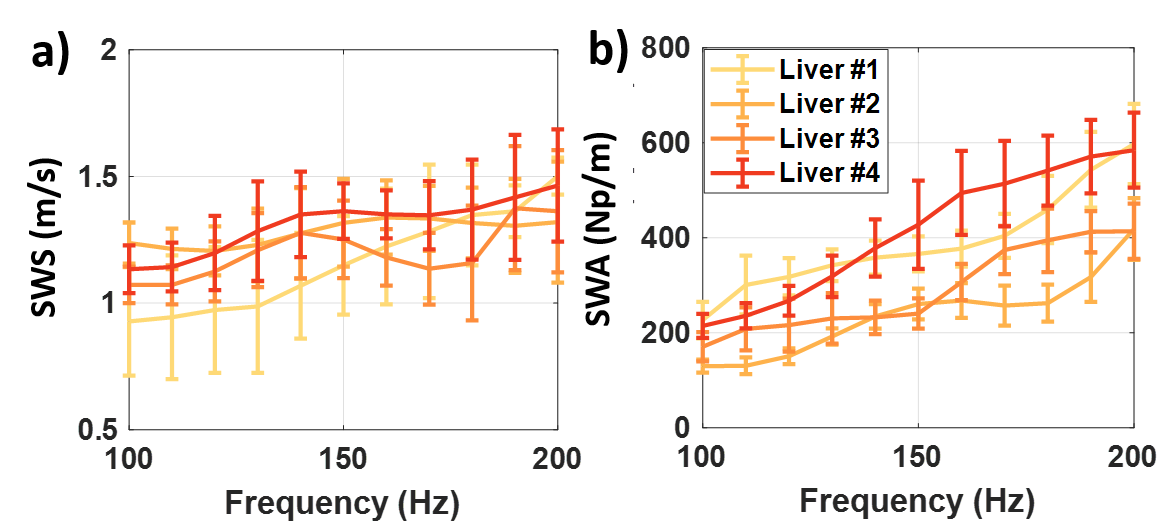}
\caption{Shear-wave phase velocity $c_p(\omega)$ (a), attenuation $\alpha(\omega)$ (b) as a function of frequency, measured for four \textit{ex-vivo} porcine livers at stress-free condition, i.e.\ $\epsilon\approx 0$.
The error bars correspond to the standard deviation of the respective estimates having R-square value greater than 0.90.}
\label{Livers_SWS_SWA_no_stress}
\end{figure}

Fig.\,\ref{Livers_SWS_SWA_no_stress} shows our estimated frequency-dependent shear-wave phase velocity and attenuation values at stress-free condition.
Livers \{\#1-\#4\}, respectively, exhibit shear-wave velocity dispersions of \{0.59, 0.12, 0.24, 0.31\},[m/s]\textbackslash 100\,Hz; and attenuation dispersions of \{305.3, 229.5, 266.0, 414.8\}\,[Np/m]\textbackslash 100\,Hz.

\begin{figure}
\includegraphics[width=\columnwidth]{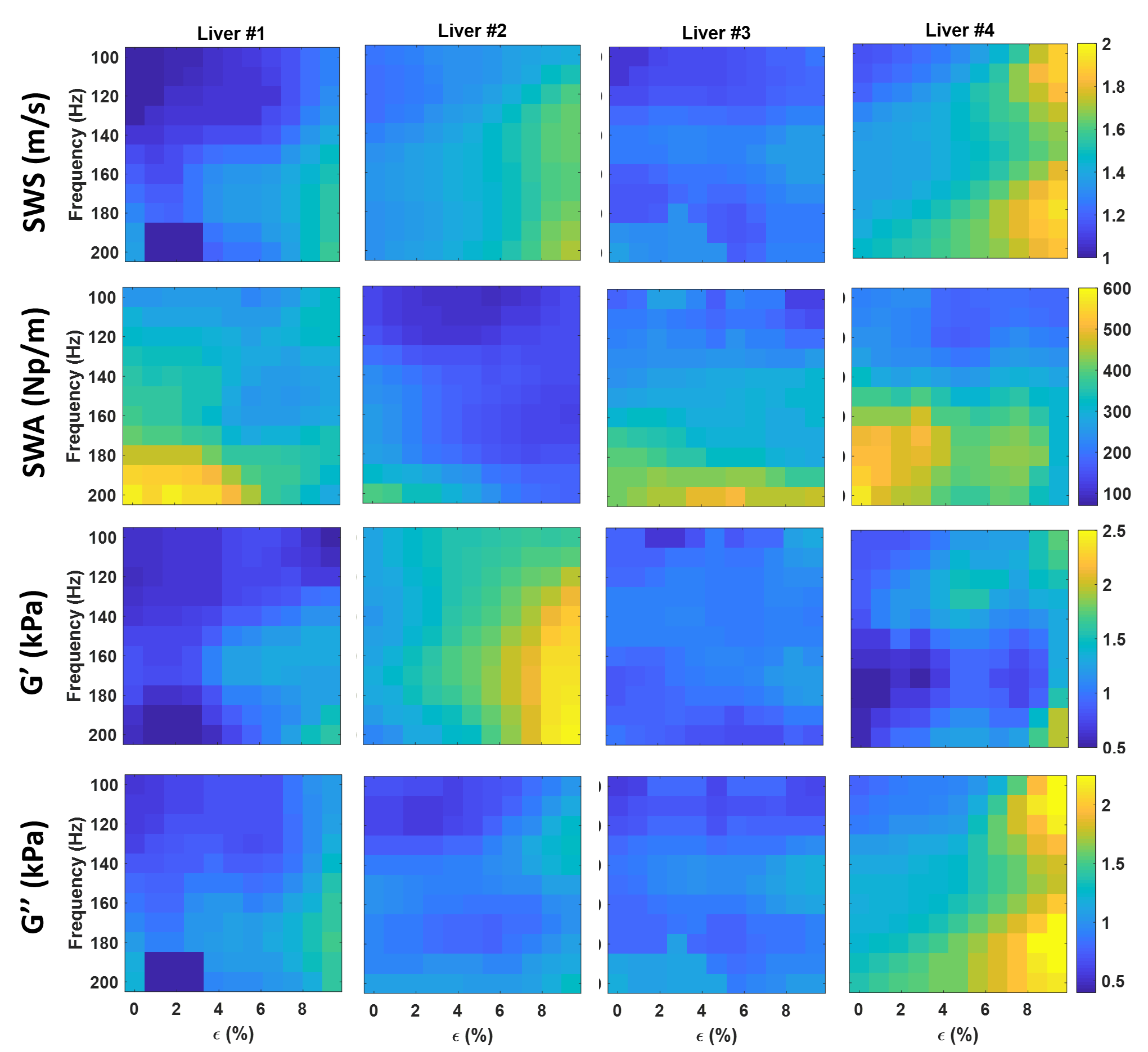}
\caption{Shear-wave phase velocity (1st row), attenuation (2nd row), storage modulus $G'$ (3rd row) and loss modulus $G''$ (bottom row) with respect to frequencies and applied strain levels for \textit{ex-vivo} porcine livers \#1 -\#4.}
\label{Livers_SWS_SWA_G_1_G_11}
\end{figure}

For these liver samples, estimated shear-wave phase velocity and attenuation, as well as the storage and loss moduli derived from the former two, are shown in Fig.\,\ref{Livers_SWS_SWA_G_1_G_11} as a function of frequencies and applied strain.
These maps are observed to be much less uniform compared to those of phantoms in Fig.\ref{Ph_SWS_SWA_G_1_G_11}, due to highly complex, nonlinear, and dispersive nature of actual tissues.

\begin{figure*}   
\includegraphics[width=\textwidth]{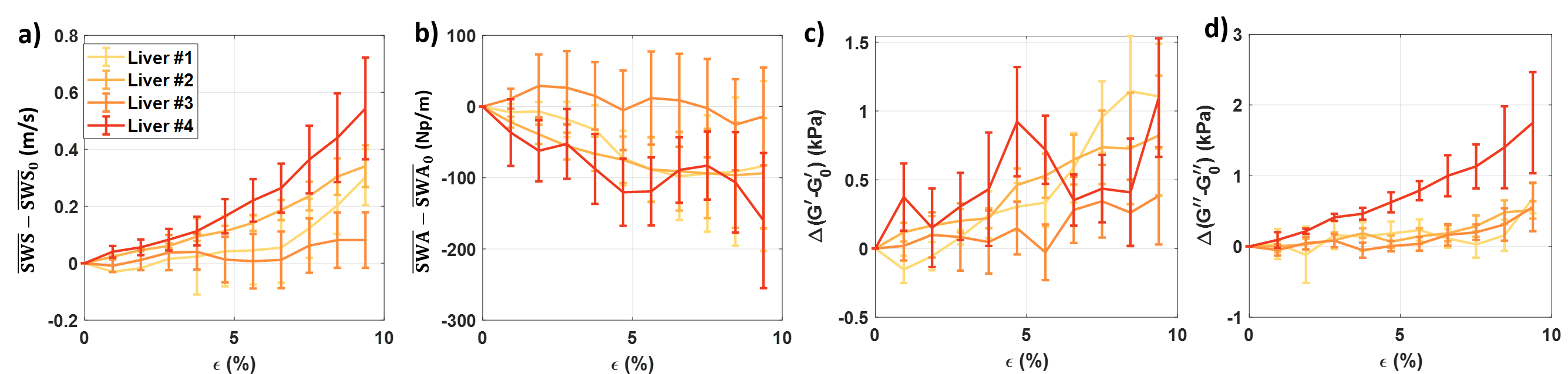}
\caption{Averaged shear-wave phase velocity (a), attenuation (b), storage (c) and loss (d) moduli with respect to frequencies on the corresponding maps in Fig.\,\ref{Livers_SWS_SWA_G_1_G_11} and offset by their nominal values, respectively, SWS\textsubscript{0}, SWA\textsubscript{0}, $G'$\textsubscript{0}, and $G''$\textsubscript{0} at stress-free state are plotted as function of applied strain ($\epsilon$), the error bars represent the standard deviation of these values.}
\label{Livers_SWS_SWA_SM_LM_wrt_compression}
\end{figure*}

Similar to the phantom experiments, shear-wave phase velocity, attenuation, storage and loss moduli were averaged with respect to frequency axis and subtracted from its frequency averaged values SWS\textsubscript{0}, SWA\textsubscript{0}, $G'$\textsubscript{0} and $G''$\textsubscript{0} at stress-free state respectively, and plotted in Fig.\,\ref{Livers_SWS_SWA_SM_LM_wrt_compression} to quantify cumulative effects in frequency with respect to applied strain. 
%and Tab.~\ref{Liver_results_Tab}
%
It is observed that, shear-wave phase velocities exhibit +17.8\%, +23.7\%, +5.6\% and +34.9\%
%0.20 (+17.8\%), 0.31 (+23.7\%), 0.07 (+5.6\%), and 0.47 (+34.9\%) 
while attenuation exhibit -28.1\%, -47.0\%, -03.7\% and -33.7\%
%-109.7 (-28.1\%), -112.0 (-47.0\%), -10.7 (-03.7\%), and -140.0 (-33.7\%) 
relative changes per 10\% applied strain in the livers \#1-\#4, respectively. 
It is interesting to note that, in contrast to phantom experiments, for the livers the changes in storage modulus with respect to applied strain resembles the changes in attenuation while changes in loss modulus with respect to applied strain resembles the changes in phase velocity.

\begin{figure}   
\includegraphics[width=\columnwidth]{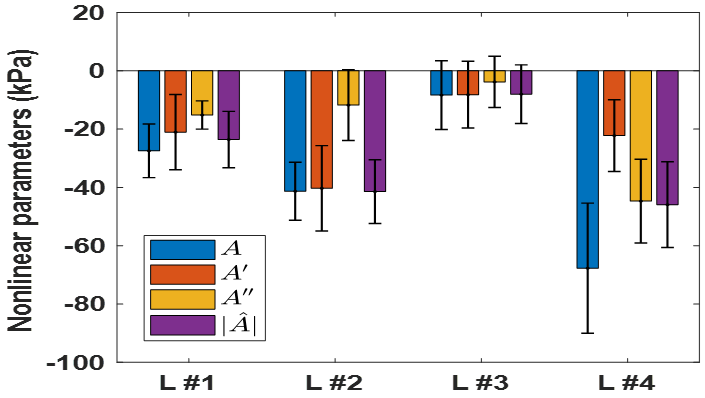}
\caption{Elastic $A$ and viscoelastic nonlinear parameters $A'$, $A''$, and its magnitude $|\hat{A}|$ are extracted from 4 \textit{ex-vivo} porcine livers that were subjected to compression of 0.3\,mm steps for a total of 3\,mm ($\approx$ 10\%). The errorbars indicate the standard deviation across frequencies.}
\label{Livers_NL_params}
\end{figure}

Fig.\,\ref{Livers_NL_params} illustrates viscoelastic nonlinear parameters $A'$, $A''$ and its magnitude $|\hat{A}|$, and nonlinear elastic parameter $A$.
These nonlinear mechanical parameters are averaged over the frequencies [100-200\,Hz] and its standard deviations are presented as error bars. 
These values also tabulated in Tab.~\ref{Liver_results_Tab}.
\begin{table*}
\caption{Measured nonlinear elastic and viscoelastic parameters ($\pm$ standard deviations) for the \textit{ex-vivo} porcine livers used in this study.}
\label{Liver_results_Tab} 
 \begin{center}
    \begin{tabular}{l||c|c||c|c||c}
     \textbf{Liver} & \textbf{$\mu_0$} & \textbf{$A$} & \textbf{$\hat{\mu}=\mu_0'+i\mu_0''$} & \textbf{$\hat{A}=A'+iA''$} & \textbf{$|\hat{A}|$}\\ 
     &\textbf{{[kPa]}} &\textbf{{[kPa]}} &\textbf{{[kPa]}} &\textbf{{[kPa]}} & \textbf{{[kPa]}}\\ 
     \hline
     \#1 & 1.05 $\pm$ 0.41 &  -27.41 $\pm$ 09.28 &  (0.58 $\pm$ 0.22) + i (0.65 $\pm$ 0.28) & (-23.80 $\pm$ 10.63) - i (15.17 $\pm$ 04.84) & -23.54 $\pm$ 09.73 \\
     \#2 & 1.55 $\pm$ 0.13 &  -41.33 $\pm$ 09.92 &  (1.20 $\pm$ 0.14) + i (0.77 $\pm$ 0.22) & (-40.31 $\pm$ 14.65) - i (11.76 $\pm$ 12.12) & -41.45 $\pm$ 10.95 \\   
     \#3 & 1.42 $\pm$ 0.23 &  -08.33 $\pm$ 11.80 &  (0.90 $\pm$ 0.15) + i (0.82 $\pm$ 0.17) & (-09.00 $\pm$ 11.83) - i (04.65 $\pm$ 09.92) & -08.03 $\pm$ 10.05 \\
     \#4 & 1.54 $\pm$ 0.36 &  -67.71 $\pm$ 22.32 &  (0.78 $\pm$ 0.23) + i (0.94 $\pm$ 0.28) & (-25.74 $\pm$ 10.59) - i (44.70 $\pm$ 14.39) & -45.92 $\pm$ 14.69 \\
   \end{tabular}
  \end{center}
\end{table*}
%Resulted nonlinear parameters shown in Fig.\,\ref{Livers_NL_params} suggests that 3rd order nonlinear viscous parameter $A''$ is proportional to nonlinear 3rd order elastic parameters $A'$ and $A$ in all the livers investigated, similar to the observations in phantom study (c.f. Fig.\,\ref{Ph_NL_params}).
Similar to the observations in the phantom experiments, the absolute value of the nonlinear elastic parameter $A$ without considering the attenuation term is +13.1\%, +2.5\%, -8.0\%, and +62.0\% higher than true nonlinear elastic parameter $A'$ measured considering the attenuation term for livers \#1-\#4.
The standard deviation of $A$ and its percentages with respect to its mean values are $\pm$33.9\%, $\pm$24.0\%, $\pm$141.6\%, and $\pm$32.9\% 
%09.28 (33.9\%), 09.92 (24.01\%), 11.80 (141.6\%), and 22.32 (32.9\%)
for livers {\#1-\#4}. This suggests a highly dispersive characteristic of actual tissue's non linear constants, similar to the observations in~\cite{otesteanu2019spectral}. 
A related trend can also be observed in nonlinear parameters $A'$ and $A''$.

%Frequency-averaged viscoelastic parameters for all the livers are presented in Tab.~\ref{Liver_results_Tab}, together with their standard deviations. 
%

%%%%%%%%%%%%%%%%%%%%%%%%%%%%%%%%%%%%%%%%%%%%%%%%%%%%%%%%%%%%%%%%%%%%%%%%%%%%%%%%%%%%%%%%%%%%%%%%%%%%
\section{Discussion}

%In this work we have studied the characteristics of acousto-elasticity effect on shear-wave attenuation and tested the feasibility of estimating nonlinear viscoelastic parameters in tissue mimicking phantoms and \textit{ex-vivo} porcine livers.
In the phantom study, an increasing castor oil percentage gave increased shear-wave velocity and attenuation dispersion profiles similar to the observations in~\cite{barry2012shear}.
As the oil percentage increases, speed-of-sound decreases and attenuation of the medium increases~\cite{nguyen2014development} which resulted in poor SNR of acquired shear-wave displacement fields.
This is evident from the larger error bars of shear-wave phase velocity and attenuation estimates for the phantom \#4 (gelatin 08\% and castor oil 15\%).
Note that the error bars of shear-wave attenuation estimates are larger than the phase velocity estimates since the amplitude spectrum profiles (Fig.\,\ref{SWS_SWA_estimation}e) of shear-waves are generally noisier than the phase profiles (in Fig.\,\ref{SWS_SWA_estimation}b).

Shear storage modulus as a function of applied strain as shown in Fig.\,\ref{Ph_SWS_SWA_wrt_compression}c suggests that nonlinearity decreases with increased castor oil percentage and is highest in the phantom with only gelatin (i.e. stiffer medium), which is in agreement with the findings in~\cite{pavan2010nonlinear}, where nonlinearity is reported to decrease with increasing oil percentage in tissue mimicking phantoms during mechanical tests. 

%Shear-wave attenuation characteristics were correlated shear-wave phase velocity characteristics as a function of applied strain that the higher the variation of phase velocity with applied strain the higher is the negative variation of attenuation as shown in Fig.\,\ref{fig:Ph_SWS_SWA_wrt_compression}a,b. (similar observations are also valid in case of \textit{ex-vivo} liver experiments.)

\begin{figure}
    \centering
    \includegraphics[width=\columnwidth]{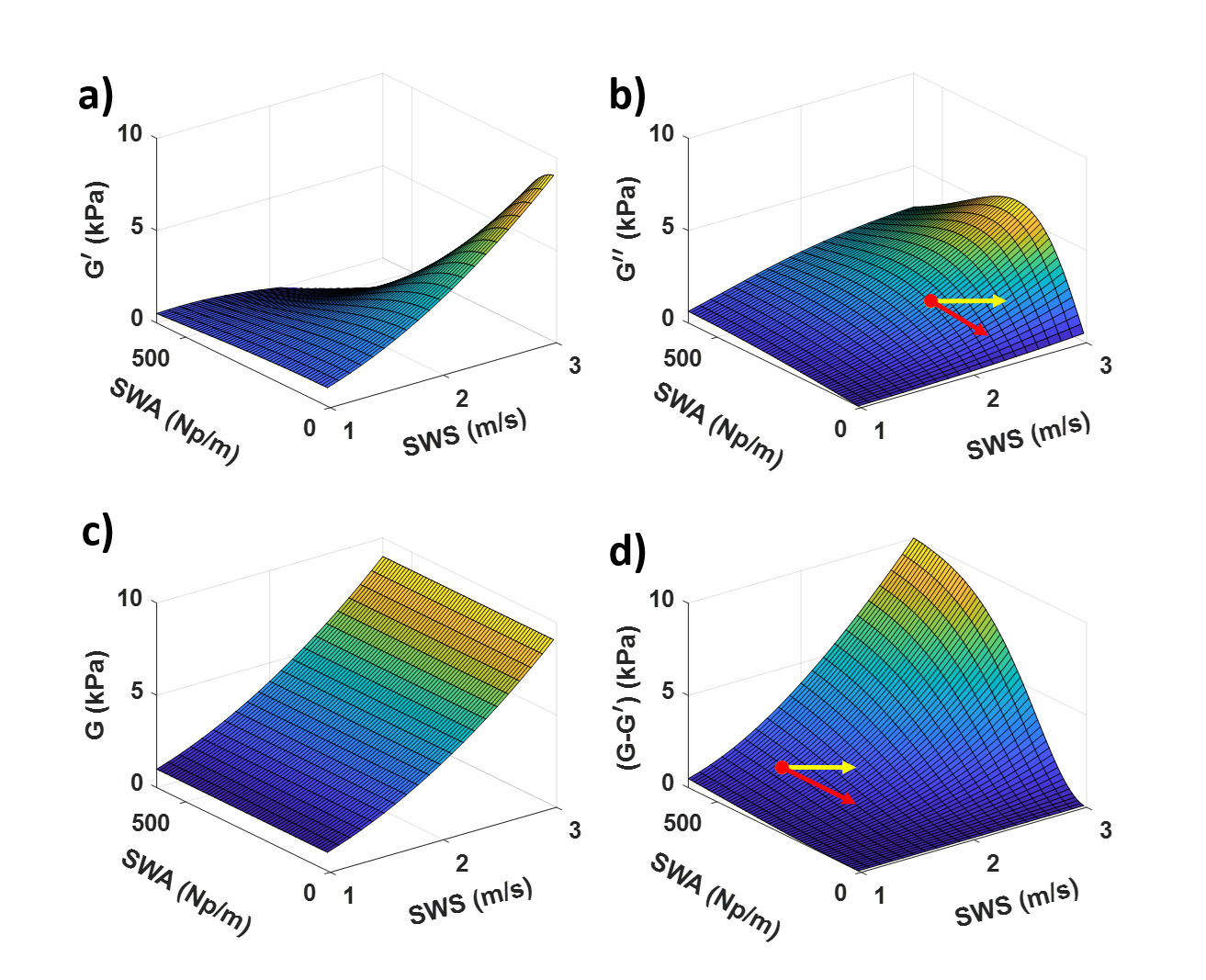}
    \caption{Storage $G'$ (a), loss $G''$ (b), and conventional shear storage $G$ (c) moduli are computed over a range of shear-wave phase velocities varying within [1,3]\,m/s and attenuation varying within [10,600]\,Np/m at frequency 200\,Hz using the empirical relationship in Eqs.\,\ref{eq:complex_mod} and \ref{eq:G} respectively. Difference between shear storage and conventional shear moduli, i.e. $G-G'$, is shown (d).
    Red and yellow arrows demonstrate, respectively, sample ascending and descending paths in (b) for $G''$ and in (c) for $G-G'$. }
    \label{fig:G_1_G_11_G_theory}
\end{figure}

It is observed that storage modulus ($G'$) was increased while loss modulus ($G''$) was decreased with applied strain as shown in Fig.\,\ref{Ph_SWS_SWA_wrt_compression}c.
Note that such $G''$ change with applied strain depends on the operating points of phase velocity and attenuation and their shifts with applied strain in a medium at a frequency.
To demonstrate this, we plotted the theoretical storage $(G')$ and loss modulus $(G'')$ values, respectively, in Fig.\,\ref{fig:G_1_G_11_G_theory}a,b 
over a range of different shear-wave phase velocity and attenuation values at 200\,Hz frequency.
Since shear-wave attenuation values decrease and phase-velocities increase as a function of applied strain, Fig.\,\ref{fig:G_1_G_11_G_theory}b suggests that a loss modulus $(G'')$ direction of variation (decrease or increase) as a function of applied strain depends on the given shear-wave attenuation and phase velocities in the phantom/tissue of interest.
For instance, if the applied strain shifts tissue properties in the direction of yellow arrow in Fig.\,\ref{fig:G_1_G_11_G_theory}b, then $G''$ would increase; whereas a change in red arrow directions would yield a reduction in $G''$.
%It can also be observed that $\Delta G'_\epsilon$, $\Delta G_\epsilon$ and $\Delta G''_\epsilon$ depends upon the range of resulted shear-wave attenuation and phase velocities as a function of applied compression, $\Delta G'_\epsilon, \Delta G_\epsilo$, and $\Delta G''_\epsilon$ are change in $G'$, $G$ and $G''$ respectively at applied strain, $\epsilon$ in our phantoms experiments and liver tissues. 
%For the range $c_p(\omega)$ and $\alpha(\omega)$ values resulted in these phantom experiments $G''$ values were seen decreased while $G'$ increased as a function of applied strain.

% \revg{It can be seen in Fig.\,\ref{Ph_NL_params} that 3rd order nonlinear parameters decreased exponentially with increasing castor oil percentage suggests the possibility of using these parameters as new bio-marker to diagnose steatosis stages in NAFLD.}
For all the phantoms used herein, $A' > A$ is observed as in Fig.\,\ref{Ph_NL_params}.
Nevertheless, this observation cannot to be generalized and depends on the material parameter change as well. %universal.
To illustrate this, the difference between shear storage modulus (Fig.\,\ref{fig:G_1_G_11_G_theory}b) and conventional shear modulus (Fig.\,\ref{fig:G_1_G_11_G_theory}c), i.e.\ $G-G'$, is plotted in Fig.\,\ref{fig:G_1_G_11_G_theory}d.
Note that the change in $G$ (or $G'$) with applied strain is the parameter $A$ (or $A'$) given Eq.(\ref{eq:all_nl_params}).
For any applied strain causing material changes along the yellow arrow in Fig.\,\ref{fig:G_1_G_11_G_theory}d, the difference \mbox{$G-G'$} is seen to increases, i.e.\ change in $G$ being larger than change in $G'$, hence $A>A'$). 
The red arrow exemplifies a behaviour opposite to the one above.

%agrees with theoretical observation in Fig.\,\ref{fig:G_1_G_11_G_theory}d that $\Delta G'_\epsilon > \Delta G_\epsilon > \Delta G''_\epsilon$ for the range of respective $c_p(\omega)$ and $\alpha(\omega)$ values.
%In the \textit{ex-vivo} porcine liver study, the shear-wave induced by acoustic radiation force push of duration 200\,$\mu$s travels limited distance due to intrinsic high shear attenuation values in soft tissues.
%Due to which, it is only possible to measure $c_p(\omega)$ and $\alpha(\omega)$ for limited frequencies [100-200\,Hz] as illustrated in Fig.\,\ref{Liver_FT_profiles}. 

%Similar to the phantom studies, $c_p(\omega)$ increased and $\alpha(\omega)$ decreased on an average $\approx$ 0.42 m/s and $\approx$ 100 Np/m respectively as a response to 10\% strain as shown in Fig.\,\ref{Livers_SWS_SWA_wrt_compression}. 
%It can be observed from Fig.\,\ref{Livers_SWS_SWA_SM_LM_wrt_compression} that the magnitude of variation in $\alpha(\omega)$ is proportional to that variation in $c_p(\omega)$, similar to the observations in phantom study (cf. Fig.\,\ref{Ph_SWS_SWA_SM_LM_wrt_compression}). 

In Fig.\,\ref{Livers_SWS_SWA_no_stress}b, attenuation is seen to increase linearly as a function of frequency, in agreement with the findings in~\cite{kijanka2019two,bernard2016frequency}.
As seen in Fig.\,\ref{Livers_SWS_SWA_SM_LM_wrt_compression}a,b, frequency-averaged shear-wave velocity and attenuation are found to be, respectively, \{+0.20, +0.30, +0.07, 0.47\}\,[m/s] and \{-109.7, -112.9, -10.7, -140.0\}\,[Np/m] for livers \#1-\#4.
In contrast, for the phantoms \#1-\#4 these values, respectively, were \{+0.08, +0.06, +0.05, $\approx$0.00\}\,[m/s] and \{-13.3, -12.9, -12.0, -1.6\}\,[Np/m] as shown in Fig.\,\ref{Ph_SWS_SWA_G_1_G_11}a,b.
Note that in both the phantom and \textit{ex-vivo} liver experiments the absolute change in shear-wave phase velocity is correlated with the absolute change in attenuation with applied strain.
This characteristic may be due the displacement of extracellular fluids and reduced gaps in extracellular matrix upon compression. This might in turn lead to a higher tissue density, thereby increasing the shear-wave velocity and decreasing the attenuation.
To study general applicability of this observation, future studies with more soft tissues are needed.

Stress-free shear moduli [1.05,1.55]\,kPa obtained in our \textit{ex-vivo} liver experiments are comparable with the values reported in~\cite{orescanin2010dispersion}, where from multiple harmonic excitations, mean shear moduli of 2.2\,kPa and 1.8\,kPa are reported, respectively, using Kelvin-Voigt and Zener models.
Average $G'$ to $G''$ ratio from our \textit{ex-vivo} liver experiments is 1.1, suggesting that the loss and storage moduli are in the same order and both need to be taken into account to fully characterize the liver or the soft tissues. 
Hence, loss modulus cannot simply be ignored and is important to consider for the AE effect to fully quantify the nonlinear viscoelastic characteristics of a soft tissue.
Third-order nonlinear elastic parameter $A$ of porcine livers were reported in~\cite{otesteanu2019spectral} in the range of [25,60]\,kPa at 200\,Hz. 
These findings are corroborated by the frequency-averaged $A$ values over [100,200]\,Hz range reported in this work.

In addition to $A'$ and $A''$, the magnitude of the third-order nonlinearity parameter, $|\hat{A}|$, reported in this study could potentially serve as a new bio-marker, along with its loss factor ($\tan \delta = A''/A'$).
For angle $\delta = 0$\textdegree{}, the medium exhibits purely elastic nonlinearity, and for 90\textdegree{} purely viscous.
Meanwhile, $|\hat{A}|$ can capture and quantify the severity of either nonlinearity.
For instance in Fig.\,\ref{Livers_NL_params}, for liver \#2,  $A'<A''$, but for liver \#4, vice versa $A'>A''$.
Magnitude of overall nonlinearity $|\hat{A}|$ is seen to be similar, while the $\delta$ changing drastically.
Such reparametrization may then provide an alternative understanding, additional correlative evidence, and potentially simpler interpretation in the clinics.

All liver samples in this study were cut into rough rectangular boxes before measurements for standardization.
Measured \textit{ex-vivo} values are likely different from their \textit{in-vivo} counterparts due to lack of perfusion and potential dehydration during the experimental process.
Note as a limitation that uni-axial compression may not be straight-forward to apply \textit{in-vivo} due to anatomical complexity and constraints around targeted tissue.
Extremities and the breast, nevertheless, can be compressed with controlled handheld setups similar to~\cite{sanabria2018speed,Rau_attenuation_19} to allow for the proposed nonlinear viscoelastic parameterization. 

%%%%%%%%%%%%%%%%%%%%%%%%%%%%%%%%%%%%%%%%%%%%%%%%%%%%%%%%%%%%%%%%%%%%%%%%%%%%%%%%%%%%%%%%%%%%%%%%%%%%

\section{Conclusion}
In this work, nonlinear characteristics of tissue viscoelasticity using acoustoelastic attenuation of shear-waves have been studied.
We first computed and studied the characteristics of frequency- and strain-dependent velocity and attenuation values of acoustic radiation force induced shear waves. Next, we derived storage and loss moduli from these values to compute nonlinear mechanical parameters by fitting parametric models of acoustoelastic theory in viscoelastic media under uniaxial compression.
We have studied these with gelatin phantoms of different oil emulsions, similarly to literature aiming to emulate NFALD.
Results indicate the following observations: ($i$)~dispersion of shear-wave attenuation and velocity increased with higher oil percentage; ($ii$)~variation of shear-wave velocity and attenuation with respect to compression decreased with higher oil percentage; and ($iii$)~magnitudes of nonlinear  viscoelastic parameters decreased with higher oil percentage. 

%From these observations, it is expected that characteristics of strain-dependent shear wave velocity and attenuation values and nonlinear visocoelastic parameters can be used for staging the liver in NAFLD disease.
%It also demonstrated that it is possible to estimate viscoelastic nonlinear parameters using the proposed method, with the dispersion and nonlinear elastic characteristics of phantoms with increased oil percentage being in agreement with those reported previously in the literature. 
A study with four \textit{ex-vivo} porcine livers were carried out to test the feasibility of the proposed methods on actual tissue.
Results reveal similar observations that the variation of shear-wave velocity is correlated with the variation of shear-wave attenuation as a function of applied strain.
%and that nonlinear viscous parameters are proportional to nonlinear elastic parameters.
With both the phantom and \textit{ex-vivo} studies, we show that using conventional shear modulus (i.e., omitting shear-wave attenuation) introduces considerable error in nonlinear parameter estimation.
Based on our results, we recommend the use of storage modulus (computed using both shear wave velocity and attenuation) for computing a true nonlinear elastic parameter. 
Nonlinear biomechanical characterization of tissues can add complementary information to conventional SWE, helping in early diagnosis and staging of different diseases such as NAFLD.
In future clinical studies our methods shall be compared with clinical gold-standard such as biopsies, to assess clinical feasibility and value.
\section*{ACKNOWLEDGMENT}
Funding provided by the Swiss National Science Foundation (SNSF) and a Swiss Government Excellence Scholarship.
%\balance
%\clearpage
\bibliographystyle{IEEEtran}
\bibliography{Bibliography.bib}
\end{document}